%% file: main.tex
\documentclass[fleqn,12pt]{wlscirep}
\usepackage[utf8]{inputenc}
\usepackage[T1]{fontenc}
\usepackage{lineno}
\usepackage{ulem}

\usepackage{tikz-cd}
\newcommand{\todo}[1]{}

\title{A Spatiotemporal Perspective on Dynamical Computation \\ in Neural Information Processing Systems\\
}

\author[1,+]{T. Anderson Keller}
\author[2,+]{Lyle Muller}
\author[3,4+]{Terrence J. Sejnowski}
\author[5,+]{Max Welling}

\affil[1]{The Kempner Institute for the Study of Natural and Artificial Intelligence at Harvard University, Cambridge, USA}
\affil[2]{Western University, Department of Mathematics, London, Ontario, Canada}
\affil[3]{Salk Institute for Biological Studies, Computational Neurobiology Laboratory, La Jolla, CA, USA}
\affil[4]{University of California at San Diego, Department of Neurobiology, La Jolla, CA, USA}
\affil[5]{University of Amsterdam, Institute for Informatics, Amsterdam, The Netherlands}
\affil[+]{All authors contributed equally to this work}


\begin{abstract}
\input{sections/0_Abstract}

\end{abstract}
\begin{document}

\flushbottom
\maketitle

\thispagestyle{empty}

\input{sections/1_Introduction}

\input{sections/3_Evidence_for_Shift}
\input{sections/5_Benefits}

\input{sections/Flow_Equivariance}

\input{sections/4_The_New_Paradigm}

\input{sections/Mechanistic_substrate}

\input{sections/Flow_Memory}

\input{sections/6_Unifying_AI_and_Neuro}

\input{sections/7_Conclusion}

\bibliography{sample}

\section*{Acknowledgements}
We would like to thank Steven Zucker, Erik Bekkers, and Arjun Karuvally for valuable discussions in the process of developing this article. T.A.K. acknowledges the Kempner Institute for the Study of Natural and Artificial Intelligence at Harvard University for funding during the process of drafting this article. L.M. acknowledges support from BrainsCAN at Western University through the Canada First Research Excellence Fund (CFREF), NSF/CIHR NeuroNex DBI-2015276, the Natural Sciences and Engineering Research Council of Canada (NSERC), the Western Academy for Advanced Research, NIH U01-NS131914, and NIH R01-EY028723. T.J.S. is supported by ONR N00014-23-1-2069 and NIH MH132664.

\section*{Author contributions statement}
All authors contributed equally to this work. 

\end{document}

%% file: sections/0_Abstract.tex
Spatiotemporal flows of neural activity, such as traveling waves, have been observed throughout the brain since the earliest recordings; yet there is still little consensus on their functional role. Recent experiments and models have linked traveling waves to visual and physical motion, but these observations have been difficult to reconcile with standard accounts of topographically organized selectivity and feedforward receptive fields. Here, we introduce a theoretical framework that formalizes and generalizes the connection between `motion' and flowing neural dynamics in the language of equivariant neural network theory. We consider `motion' not only in physical or visual spaces, but also in more abstract representational spaces, and we argue that recurrent traveling-wave-like dynamics are not just useful but necessary for accurate and stable processing of any signal undergoing such motion. Formally, we show that for any non-trivial recurrent neural network to process a sequence undergoing a flow transformation (such as visual motion) in a structured equivariant manner, its hidden state dynamics must actively realize a homomorphic representation of the same flow through recurrent connectivity. In this ``spatiotemporal perspective on dynamical computation'', traveling waves and related flows are best understood as faithful dynamic representations of stimulus flows; and consequently the natural inclination of biological systems towards such dynamics may be viewed as an innate inductive bias towards efficiency and generalization in the spatiotemporally-structured dynamical world they inhabit.

%% file: sections/1_Introduction.tex
\section*{Introduction}

For much of modern neuroscience, cortical computation has been cast through the lens of filters: hierarchies of neurons that extract increasingly abstract features from the sensory stream. From the early discoveries of retinal receptive fields\cite{retinotopy, Kuffler} and the simple/complex cells of Hubel \& Wiesel \cite{hubel_wiesel_rf}, to the face\cite{ffa}, place\cite{ppa}, and body\cite{eba} patches of inferotemporal cortex, the prevailing intuition has been that perception is achieved by a progressive cascade of filtration and combination, abstracting stimulus space into a selective mosaic of neural detectors. This feature-centric paradigm, rooted in early single-unit recordings and later formalized in the Parallel Distributed Processing movement \cite{Rumelhart1986-nl}, has proved enormously productive, seeding both classical theories of perception and today’s artificial intelligence\cite{HMAX, Rosenblatt1958, Fukushima1980, DeepLearning2015}. Yet, in the last few decades, as neural recording technologies have continued to develop, evidence has accumulated which has come into tension with this account. 

Advances in multi-electrode arrays, voltage-sensitive dyes, and wide-field calcium imaging have provided an aerial view of population activity, revealing a portrait painted not in static receptive fields, but in waves, spirals, and traveling fronts of neural activity that sweep across cortex\cite{hughes1995, ERMENTROUT200133, Sato2012, MULLER2012222, Muller2018}. Sensory evoked responses seldom remain confined to the crystalline tiling implied by selectivity; instead, they expand and propagate, giving rise to richly structured spatiotemporal dynamics that are difficult to reconcile with a purely feed‑forward, feature‑detector account. If a neuron fires because a traveling wave happens to pass over its cortical territory, what does its ``preferred stimulus'' really mean?

We propose that this apparent conflict dissolves once one shifts perspective. Rather than asking what neurons are selective for, we ask what representations an embodied agent needs in order to act coherently in a world that unfolds over space and time. In natural settings, stimuli do not arrive as independent frames, they arrive as a continuous stream, smoothly transforming from one instant to the next. The computational problem, then, is not merely to recognize static features, but to maintain \emph{structured} internal variables as inputs move: to preserve identity under motion, to predict lawful transformation, and to support inference and control without having to relearn the same content in every moving frame of reference.

Here we propose a spatiotemporal perspective that not only alleviates the tension between traveling waves and structured neural selectivity, but also provides a mandate for local recurrent spatiotemporal dynamics -- to shift representations in step with the movement of the input so that the same feedforward computations can be reused as the world flows. In other words, spatiotemporal flows of neural activity can implement the internal ``change of reference frame'' that keeps computation stable across moving vs. stationary stimuli, allowing embodied networks to operate in effectively co-moving frames with the constantly shifting environments they inhabit. 

We formalize this idea through \emph{flow equivariance}. In the language of geometric deep learning, the smooth transformations that dominate natural experience can be understood as flows: time-parameterized symmetries capturing structured change over time. Flow equivariance is then the requirement that a recurrent neural network must process moving inputs in a way that is identical to how it would process stationary inputs, but with equivalent movement applied to its hidden state. 
Crucially, our theory asserts that for any non-trivial recurrent network to achieve such flow equivariance, its hidden state dynamics must  \emph{actively flow} in a manner homomorphic to the input flow. Put differently: if the world moves, the latent state must preemptively move in unison through recurrent dynamics. In this perspective, we will generalize the idea of `motion' to flow in an abstract representation space, and demonstrate how such flows can additionally be seen to serve as a form of memory.
We further argue that once one takes biological constraints such as conduction velocity and metabolic costs seriously, the most natural realization of these latent flows is a local spatiotemporal flow -- traveling-wave-like dynamics supported by lateral connectivity. %

In the following, we will outline the mounting evidence for this perspective, empirically linking spatiotemporal neural dynamics and movement throughout neuroscience literature. We will describe from a theoretical perspective why the `structure of movement' is best represented as a flow; and further provide an argument supporting why that flow would manifest as clear spatiotemporal dynamics in biologically constrained neural networks. We will then argue that the natural bias of biological neural networks towards such structured spatiotemporal dynamics may be best seen as an inductive bias towards properly representing the structured transformations of the world around us. We will conclude with a review of evidence from both modern artificial intelligence and experimental neuroscience which supports this perspective, and finish with a number of predictions that our generalized perspective supports. 
Ultimately, our goal is not to discard the ideas of selectivity, but to embed them within a broader spatiotemporal perspective that acknowledges that brains, like the worlds they inhabit, are spatially embedded dynamical systems, and these spatiotemporal dynamics themselves should be considered foundational.

%% file: sections/3_Evidence_for_shift.tex
\section*{Spatiotemporal Dynamics in Neural Systems}

The study of spatiotemporal dynamics in neural activity dates back to the origins of neuroscience itself. In the following, we will summarize a portion of this history and emphasize the historical pervasiveness of such observations. We refer readers to the review articles of Hughes (1995)\cite{hughes1995}, Ermentrout and Kleinfeld, (2001) \cite{ERMENTROUT200133}, Sato et al. (2012)\cite{Sato2012}, and Muller et al. (2012 \& 2018)\cite{MULLER2012222, Muller2018} for a comprehensive overview of this history. We will follow this historical discussion with a review of the more recent literature on spatiotemporal dynamics, highlighting a significant portion of work that appears to align such dynamics with `motion', both visual and physical.

\input{sections/3.1_Experimental_Evidence}

\input{sections/3.2_Computational_Models}

%% file: sections/3.1_Experimental_Evidence.tex
\subsubsection*{From the Earliest Neural Recordings}
The first known recorded evidence of traveling waves of neural activity was reported by Adrian and Matthews in 1934. The authors reported the discovery of an oscillatory `injury response' in electrode recordings measured across the cortex of anesthetized rabbits and cats, stating ``a single pulsation may travel like a ripple over the surface of the brain", further noting that "an almost identical pattern of activity may appear at points 10 mm or more apart"\cite{adrian1934}. These initial observations were followed by a significant number of EEG studies which measured systematic offsets of the phase of neural oscillations (primarily alpha oscillations) with respect to spatial location on the cortex, and hypothesized that traveling waves were the root cause\cite{adrian1934_2, cohn1948, Brazier1949}. Some of the first studies more precisely characterizing traveling waves of neural activity were performed by Goldman et al. (1948 \& 1949)\cite{goldman1948, goldman1949} and Lilly (1949)\cite{lilly1949method}, through the introduction of 16-element and 25-element electrode arrays respectively, producing the first recordings of what Lilly terms `electro-iconograms' from animals. In the isolated visual cortex of cats, Burns (1950 \& 1951) \cite{burns1950some, burns1951some}, subsequently measured the velocity of traveling waves of activity; and Mickle and Ades (1953) recorded some of the first clear traces of propagating cortical traveling waves in the cat auditory cortex evoked from audible clicks. In the words of Mickle and Ades: "The nature of the recording device has rendered interpretation difficult [...] There is no doubt, however, that wave forms appear to spread across the cortical surface"\cite{mickle1953}.  Unfortunately, however, since traveling waves are difficult to detect in recordings from single neurons, these pioneering studies were subsequently eclipsed when single-unit recordings from cortical neurons became dominant in the 1960s. 

\begin{figure}
\centering
\fbox{
\begin{minipage}{0.99\textwidth}
\centering
\vspace{1mm}
\textbf{\color{brown} Box 1. Traveling Waves in Natural and Artificial Neural Systems 
}
\vspace{1.5mm}
\hrule
\vspace{1.5mm}
\includegraphics[width=0.98\textwidth]{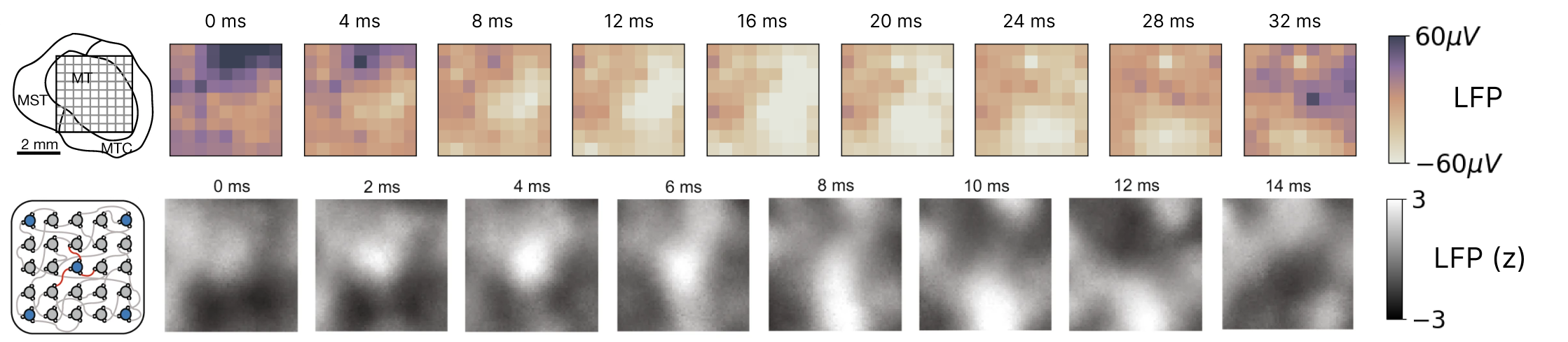}
\captionsetup{labelformat=empty}
\caption{\textbf{Recording of traveling waves from area MT of an awake behaving primate and simulations in a spiking neural network model.} Top: Array of electrodes reveals structured spatiotemporal dynamics (adapted from ref.\cite{Davis2020}). Bottom: a simulated spiking neural network model with topographically constrained connectivity and distance-dependent time delays (adapted from ref.\cite{Davis2021}). The dynamics in both cases are similarly structured as local, spatiotemporally structured flows in space and time. }
\label{fig:recorded_waves}
\end{minipage}
}
\end{figure}

\subsubsection*{New Recording Technology; Renewed Interest}
Over time, with the continued development of multi-electrode recording devices, voltage-sensitive dyes, and optical imaging, our ability to measure such structured spatiotemporal activity has expanded dramatically, leading to a renewed interest in these dynamics. For example, traveling waves have now been observed and measured directly in the visual \cite{Grinvald2545, cat_visual_waves}, auditory \cite{auditory_waves}, somatosensory \cite{sensorymotorwaves_mouse, gonzales2025}, and motor \cite{motor_waves} cortices across a range of states of awareness and stimulation conditions\cite{Davis2020, Muller2016, Muller2014, Muller2018}. Wave dynamics have similarly been observed across nearly all scales and spatial locations in the brain, including theta, alpha, and beta oscillations in the cortex \cite{Zhang2018}, hippocampus \cite{Lubenov2009} and thalamus \cite{MULLER2012222, kim1995}; and influential studies have been published demonstrating that large scale neural activity can often best be described in terms of idealized waves traveling over the spatial geometry of the cortex \cite{NUNEZ1974279, pang, Leech2024.06.17.599309}. Through precise experimental controls and training, traveling wave-like dynamics have been directly demonstrated to have behavioral consequences:  waves in the visual cortex of non-human primates have been shown to gate perception of low-contrast stimuli\cite{Davis2020} (Box \ref{fig:recorded_waves}, top); feedforward theta waves linking visual and frontal cortex have been demonstrated to predict behavioral reaction times\cite{Luo2024}; and theta-band traveling waves in the frontal eye fields have been shown to rhythmically modulate both accuracy and reaction time in a primate visual working-memory change-detection task\cite{Han2026}. The diversity of these observations has naturally led to a diversity of hypothesized roles for wave-like dynamics, including: the transfer of information \cite{besserve_2015}, the consolidation of long term memories during sleep \cite{Muller2016}, the modulation of visual processing \cite{waves_modulate_vision, Townsend10074}, the sequencing of cortical regions \cite{region_sequencing}, the structuring of cortical space \cite{Sato2012}, the encoding of motion \cite{motionwaves}, the storage of short-term \cite{King2021} and working memory \cite{working-memory-waves, miller_spatial_computing, Whittington2023.11.05.565662, WM_NAS}, and  more\cite{Raut2021}.

\subsubsection*{A Link between Waves and Visual Movement}
Despite the apparently incongruent set of publicized roles for spatiotemporal neural dynamics in both experiment and theory, one through-line has recently emerged. Namely, a large number of results appear to link spatiotemporal neural dynamics with `movement' in both physical and visual spaces. In visual space in particular, traveling waves have been measured to correlate with the motion of stimuli across the visual field in a variety of different manners. For example, electrode-array recordings in primate area MT show that drifting gratings evoke traveling waves whose single-trial propagation direction is sensitive to the stimulus motion direction~\cite{Townsend10074}.
Complementarily, recent neural-field models demonstrate a concrete mechanism by which intrinsic traveling waves shaped by lateral connectivity can resonate with moving stimuli to yield direction and speed-selective responses~\cite{motionwaves}. 
Traveling waves have also been demonstrated to track perceived motion, not just physical motion: illusory-motion stimuli evoke wave patterns that closely match those elicited by true motion, whereas visually similar control stimuli that do not induce the illusion fail to produce the same wave dynamics \cite{jancke, Chemla4282}. 
In more abstract modeling work, it has been shown that simple networks with intrinsic wave dynamics in their recurrent spaces significantly outperform networks with unstructured dynamics at predicting the future of simple video sequences with object motion~\cite{Benigno2022}. When this wave-like structure is not built into models a priori, recent work has shown that simply pre-training networks with wavelike stimuli (simulating retinal waves before eye opening) drastically improves modeling performance of video sequences collected from agents undergoing egomotion~\cite{may2025}. Finally, state-of-the-art work in neural field theory has demonstrated analytically that stimulus motion can entrain waves in neural fields in order to accurately track stimulus spatial location with propagating wavefronts~\cite{waves_motion_adaptive_fields}. 

It is interesting to contextualize these new findings with respect to well known, canonical circuitry for movement detection. Specifically, directional `motion' selective cells have been found in the cortex of cats as far back as 1959 by David Hubel \cite{hubel_movement}, quickly followed by many similar discoveries in the optic tectum of frogs and pigeons, and in the retinae of rabbits \cite{MaturanaLettvinMcCullochPitts1960JGenPhysiol, MaturanaFrenk1963Science, BarlowHillLevick1964JPhysiol}.
Beyond early visual areas, higher level cortical areas such as area MT have been extensively
characterized in terms of motion selectivity. Early primate recordings established MT as a
motion-specialized region whose neurons respond strongly to moving stimuli and are sharply tuned for
both direction and speed \cite{Zeki1974FunctionalOrganizationMT,MaunsellVanEssen1983MTProperties}. Similar to the orientation columns of primary visual cortex, MT has been found to be organized in a columnar structure with topographic organization of direction selectivity \cite{AlbrightDesimoneGross1984MTColumns,BornBradley2005MTReview}.

At the algorithmic level, many canonical models of direction selectivity can be understood as comparing
signals across space at different times. Classic correlator models (e.g. the Hassenstein--Reichardt
detector) achieve this by combining a spatial offset with an explicit delay-and-compare computation
\cite{hassenstein1956systemtheoretische}. A closely related and highly influential viewpoint is the
spatiotemporal energy model, which casts direction and speed tuning as the output of linear filters
oriented in space-time, followed by simple nonlinearities and opponency
\cite{AdelsonBergen1985MotionEnergy}. These energy computations also form a bridge
between V1-like direction-selective subunits and MT-like velocity tuning via pooling and normalization
\cite{SimoncelliHeeger1998MTModel}.

Crucially for our purposes, the computational ingredients that generate direction selectivity -- spatially structured coupling on a retinotopic map together with biophysically constrained delays -- are the same ingredients that generate propagating activity in spatially extended recurrent networks. In other words, the ``delay-and-transport'' computation that detects movement at the circuit level can be viewed as a local manifestation of a transport operator that, at the population level, appears as a traveling wave. In the remainder of this perspective we will make this link precise: when we formally introduce flow equivariance, we will show how classic direction-selective motion computations can be interpreted as simple instances of flow equivariant computation at the multi-cellular circuit level.

\subsubsection*{Non-Visual Movement: Flows}

Beyond vision, the relationship between traveling waves and movement is not confined to visual stimuli. With respect to physical movement, beta oscillations have been observed to propagate across macaque motor and premotor cortex along dominant axes \cite{motor_waves}, and similar propagating beta waves have been reported in human primary motor cortex \cite{Takahashi2011}. These propagating dynamics additionally appear behaviorally relevant: patterned intracortical microstimulation that counter-propagates the endogenous wavefronts delays reaction time, providing causal evidence that propagating excitability patterns participate in movement initiation \cite{Balasubramanian2020}; and related work has characterized rich transient beta-LFP wave patterns during preparatory steady states \cite{Rule2018}. At the spiking level, large-scale spatiotemporal spike patterning has been shown to align with beta-band wave propagation, with task-relevant sequential spiking strongest for neuron pairs oriented along the dominant wave axis \cite{Takahashi2015}. Beyond cortex, it is also well known that hippocampal theta frequency and power increase with running speed \cite{Kennedy4326, geisler2007}. In models, wave-like dynamics have long been implicated as a core ingredient of spinal central pattern generators \cite{waves_spinal}; consistent with this picture, multisegment recordings in cat spinal cord during fictive scratching reveal rostrocaudally propagating sinusoidal field-potential waves whose cycle duration is strongly correlated with scratch cycle duration, building on earlier reports of scratch-related propagating cord-dorsum potentials \cite{BayevKostyuk1981}.

Additionally there is some evidence linking areas that are known to process visual motion with motion in more abstract spaces. 
For example, it has been demonstrated that area MT/V5 shows activation for abstract auditory `motion' \cite{POIRIER2005650, Saenz5141}. 
It is also known that certain patients with symptoms of cerebral akinetopsia (loss of motion vision) appear to lose a more general ability to integrate sequential information (motion through abstract space) into coherent wholes. For example, in 2003, a 60-year-old man who reported an inability to see motion after a traumatic brain injury also reported: "losing track of who is talking during group conversations; losing his place while visually scanning a written document in the vertical and horizontal directions; and an inability to visualize intended three-dimensional images from the two-dimensional representations on a blueprint" \cite{Pelak01012005}. 
While these examples are limited in number, they are suggestive that perhaps there is at least some shared mechanism between visual and non-visual movement processing.

\begin{figure}[h!]
\fbox{%
\begin{minipage}{0.99\textwidth}
\begin{center}
\textbf{\color{brown} Box 2. Generalized Movement as Flow} \vspace{-3mm} 
\end{center}
\hrule 
\vspace{1mm}
\hspace{0.1cm}
\centering
\includegraphics[width=1.0\textwidth]{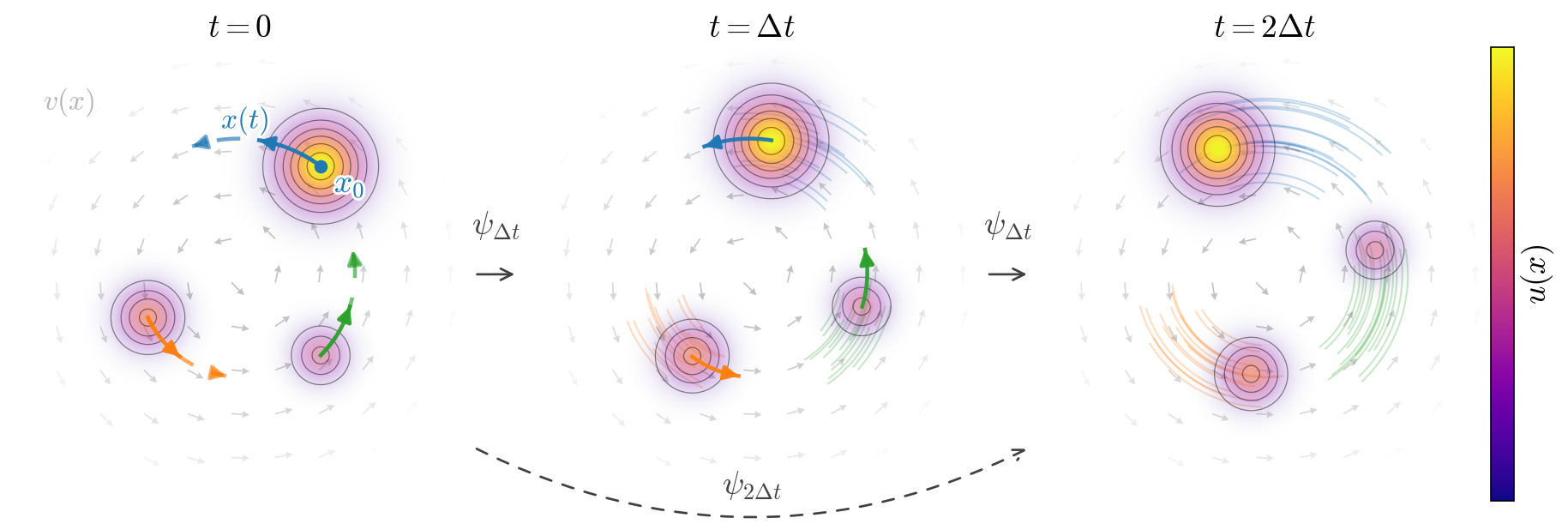}
\captionsetup{labelformat=empty}
\caption{\textbf{Generalized structured `movement' over time can be formalized as a \emph{flow} of a signal $u(x)$.} Formally, a \emph{flow} is a continuous family of transformations $\{\psi_t\}_{t\in\mathbb{R}}$ on a space $X$ (e.g., pixel coordinates, joint-angle configurations, or a neural activity manifold). Given a vector field $\nu(x)$, trajectories follow the differential equation $\dot{x}=\nu(x)$, and the flow map $\psi_t$ sends an initial condition $x_0$ to its evolved state $x(t)=\psi^{(\nu)}_t(x_0)$ after following $\nu$ for $t$ timesteps. The defining property of a flow is the \emph{one-parameter group law}: $\psi_0=\mathrm{Id}$, $\psi_{t+s}=\psi_t\circ\psi_s$, and $\psi_{-t}=\psi_t^{-1}$, reflecting the additive structure of time. In this way, a flow $\psi_t$ can be thought of informally as a time-indexed group element $g_t$. A flow can act on scalar (or channel-wise) signals defined over space via the pullback action $(\psi_t\cdot u)(x)=u(\psi_{-t}(x))$, which recovers familiar movement operations such as translation $(\psi_t^{(\nu)}\cdot u)(x)=u(x-\nu t)$ when $\psi_t^{(\nu)}(x)=x+\nu t$ (for fields with nontrivial transformation of values, an additional representation factor appears). In this perspective, `motion' in pixels, posture, or latent neural coordinates is unified as a flow -- making group structure and equivariance natural targets for both biological dynamics and engineered architectures.
}%
\label{fig:flow_vis}%
\end{minipage}}
\end{figure}

Together, this evidence points towards a connection between wave-like neural dynamics and a generalized form of `motion'. In the following sections, we will provide a mathematical formalism demonstrating that this connection is not merely by chance, but rather that waves are necessary for the accurate representation of such motion. To do so, we will think of generalized `movement' in abstract space as a \emph{flow} acting on a signal beyond just the physical or visual environment (Box \ref{fig:flow_vis}). 
For example, visual movement can be seen as the flow of pixel values across an image; physical movement of one's body can be seen as a flow acting on the space around oneself, propagating exactly in the direction opposing self-motion; and auditory movement can be seen as a flow acting on a signal defined over frequency space.

By abstracting `movement' to general flow-based transformations, we can leverage the mathematical structure of \emph{groups} -- sets of transformations that jointly obey a precisely structured algebra. This structure is what allows developers of \emph{group equivariant neural networks} to embed the symmetries that these groups define directly into network architectures, and thereby gain improved data efficiency and generalization properties for data that contains equivalent symmetries. As we will outline through this perspective, traveling waves of neural activity are the natural choice for integrating the group structure of `movement' into biological neural systems. We therefore believe encouraging this abstract group structure in neural dynamics, and reaping its associated efficiency benefits, may be one of the primary normative reasons why neural networks are innately biased towards the production of spatiotemporal dynamics. 

%% file: sections/5_Benefits.tex
\input{sections/5.1_Benefits_for_ML}

%% file: sections/5.1_Benefits_for_ML.tex
\section*{Representation Learning in a Dynamic Environment}
Before formalizing the notion of equivariance, it is helpful to take a step back and understand more generally what makes a `good' representation. 
Consider a natural image -- a full megapixel array representing an image is very high dimensional, but the parts of the image that need to be extracted are generated by a much lower-dimensional process. For example, imagine a puppet that is controlled by nine strings (one to each leg, one to each hand, one to each shoulder, one to each ear for head movements, and one to the base of the spine for bowing); the state of the puppet could be transmitted to another location by the time course of 9 parameters, which could be reconstructed in another puppet. It is therefore more efficient to represent the world in terms of these lower-dimensional factors in order to be able to reduce the correlated structural redundancies in the very high-dimensional data. %
Accurately compressing these correlations over space and time is fundamental to representing the world efficiently and enabling action in a complex environment; furthermore, generalization would be limited if every image were stored like cameras. This was the fate of Funes el Memorioso \cite{borges1944funes}, a fictional character from Argentine writer Jorge Luis Borges. In that story the character had an accident and could remember every single detail. As a consequence, he was no longer able to generalize -- he failed to capture any correlation structure from one scene to the next and ended up unable to live a normal life. Although this story is fictional, there exists a very real counterpart, chronicled by Aleksandr Luria \cite{Luria1986-my}, highlighting that the core concept is not only a theoretical idea but a practical concern for real world learning systems.

\subsubsection*{The Goal of Learning -- Structured Abstraction}
At a high level, one can understand the goal of `learning' with deep neural networks as attempting to construct these useful factors that are abstractions of the high-resolution degrees of freedom in sensory inputs, exactly like inferring the control strings of the universal puppet master. This is similar to the way that the renormalization group procedure in physics compresses irrelevant degrees of freedom by coarse graining to reveal new physical regularities at larger spatiotemporal scales, and is a model for how new laws emerge at different spatial and temporal scales. As agents in the complex natural world, we seek to represent our surroundings in terms of useful abstract concepts that can help us predict and manipulate the world to enhance our survival. %

One early idealized view of how the brain might be computing useful abstract representations is by learning features that are \emph{invariant} to a variety of natural transformations. This view was motivated by the clear ability of humans to rapidly recognize objects despite their diverse appearance at the pixel level while undergoing a variety of natural geometric transformations, and further by the early findings of Gross et al. (1972)\cite{gross} that individual neurons in higher levels of the visual hierarchy (inferior temporal cortex, IT) responded selectively to specific objects irrespective of their position, size, and orientation. %

However, even early on, many visionary researchers began to show interest in the idea that perhaps the optimal way to achieve the desired compression and invariance was not by throwing out all natural transformations, but instead by explicitly learning the symmetries of the world in a generalizable manner and selectively `undoing' them to yield more stable representations. For example, William Hoffman (1966 \& 1968)\cite{william_c__hoffman_1966, hoffman1968neuron} was one of the first to explicitly formulate how the brain might be accomplishing this, by comparing specific symmetry transformation generators (Lie derivatives) with individual cell morphologies in visual cortex; yet others have attributed the general idea as far back as Akishige (1961)\cite{akishige1961theoretical}, Rosenblatt (1960)\cite{rosenblatt1960perceptual}, McCulloch and Pitts (1947)\cite{pitts_equivariance}, and Cassirer (1944) \cite{cassirer1944concept}.

\subsubsection*{Equivariance: Building Symmetry Structure into Neural Networks}
In the machine learning literature, the idea of structurally representing transformations of features has been described under a variety of umbrella terms including: untangled `object manifolds'\cite{DICARLO2007333}, `disentangled' representations\cite{higgins2018definitiondisentangledrepresentations}, `capsule networks'\cite{NIPS2017_2cad8fa4}, and `equivariant neural networks'\cite{cohen2016group}. Although these frameworks differ in their details, one commonality they share is that an object's identity is defined by its set of symmetry transformations, and these transformations should therefore be structurally embedded in the models themselves\cite{pmlr-v32-cohen14, cohen2015transformation, higgins2018definitiondisentangledrepresentations}. 

\begin{figure}[h!]
\fbox{%
\begin{minipage}{0.99\textwidth}
\begin{center}
\textbf{\color{brown} Box 3. Equivariant Convolutional Neural Networks} \vspace{-3mm} 
\end{center}
\hrule
\vspace{1mm}
\hspace{0.1cm}
\includegraphics[width=0.98\textwidth]{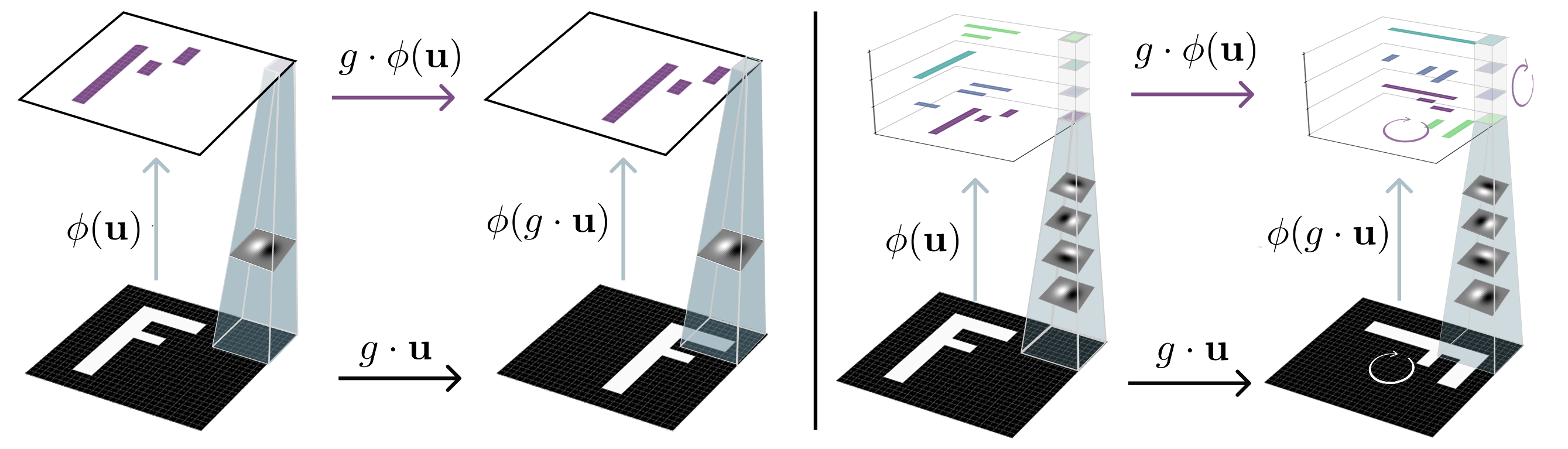}
\captionsetup{labelformat=empty}
\caption{\textbf{Equivariant feature extractors $\phi(\mathbf{u})$ are those that commute with a given transformation of interest and, in doing so, they \emph{preserve the structure of the space they act on}}. In this way, the notion of space is expanded to include the space of arbitrary transformations and one can naturally describe `motion' in that space, just like moving in physical space.\\ \vspace{1.5mm}
In the left figure above, we depict a convolutional feature extractor $\phi$ applying an edge detection kernel to a simple image $\mathbf{u}$. We see that regardless of whether we first move the object in image (through the spatial translation operation $g \cdot \mathbf{u}$) and then convolve ($\phi(g \cdot \mathbf{u})$), or if we first convolve the image $\phi(\mathbf{u})$ and then translate the output, the result is unchanged, i.e. $\phi(g \cdot \mathbf{u}) = g \cdot \phi(\mathbf{u})$. This is due to the equivariance of the convolution operation with respect to the translation group. However, traditional convolutional neural networks are not by default equivariant with respect to the group of 90-degree rotations. In the right figure, we depict a rotation-equivariant convolution\cite{cohen2016group} where the output now exists in a `lifted'  rotation space (the vertical axis), constructed by convolving the input with rotated copies of the original kernel. We see in this setting, when rotation is applied to the input, the output transformation takes the structured form of cyclically permuting the activations through the lifted space combined with the usual spatial rotation. The model thus has a \emph{structured} latent space with respect to translation and rotation.
}
\label{fig:commuting_diagram-a}
\end{minipage}%
}
\end{figure}

Equivariant neural networks in particular are built to explicitly respect known symmetries of the input in an analytically constrained manner. Simply, an equivariant representation is one where the output of our neural network transforms in a predictable way under transformations of the input. For example, we want a segmentation mask to rotate if we rotate the input image. More formally, equivariance can be expressed by the requirement that first transforming the input $\mathbf{u}$ by a transformation $g$ and then mapping through a neural network layer $\phi$, should give the same result as first mapping and then transforming: $\phi(g \cdot \mathbf{u}) = g \cdot \phi(\mathbf{u})$ (Box \ref{fig:commuting_diagram-a}).
This is called a homomorphic representation, a structure-preserving map between two spaces. We note that in general, the transformation $g$ may act on the input and output spaces differently, and therefore the equivariance relation is sometimes written as $\phi(T_g [\mathbf{u}]) = T'_g[ \phi(\mathbf{u})]$. In the remainder of this work, we use the more compact $(g \ \cdot )$ action notation for conceptual simplicity.

Equivariant representations have been shown to be beneficial for both reducing the number of required training examples (sample complexity)\cite{winkels2018d} and improving test performance (generalization)\cite{cohen2016group}. The most ubiquitous and well known equivariant neural network architecture is the convolutional neural network, equivariant to spatial translations, which has become dominant for image modeling precisely because of its improved data efficiency and generalization properties. As a further testament to its benefits, equivariance with respect to known physical symmetries is now a cornerstone of state of the art applications of deep neural networks to scientific data \cite{alphafold, pmlr-v119-bogatskiy20a, hoogeboom2022equivariantdiffusionmoleculegeneration}. It has even been further argued that accurately representing symmetries and their associated invariant quantities is the most powerful form of generalization possible in machine learning\cite{higgins2018definitiondisentangledrepresentations}. 

\subsubsection*{The Structure of Motion}
To date, mirroring the dominant `filter' paradigm of thought in neuroscience, the major focus in machine learning research has been on feedforward neural networks such as convolutional neural networks and transformers. In the subfield of equivariant deep learning, the research frontier has therefore similarly largely focused on the `static' symmetry transformations that affect the input to these models. %

However, embodied experience in the natural world is clearly dominated by motion transformations. From the swaying of trees in the wind, to the visual flows induced by self-motion, instantaneous moments of our experience are almost necessarily interrelated by smooth continuous motion transformations over time. Recently, equivariant deep learning theory has been extended to encompass recurrent neural networks and these time-parameterized transformations through the framework of `flow equivariance'\cite{fernn}. At a high level, traditional equivariance  describes how to build feedforward neural networks such that their latent space accurately reflects the structure of the instantaneous snapshots of the world. Flow equivariance ties these instantaneous moments together, describing how to construct latent \emph{dynamics} such that they accurately reflect the structure of the dynamics of the world.
In the following we describe flow equivariance more formally, and suggest that spatiotemporal neural dynamics are the natural candidate to induce this kind of dynamic structure.%

%% file: sections/Flow_Equivariance.tex
\section*{Flow Equivariance: Properly Representing Motion Requires Latent Flows}
\label{sec:flow_eq}
We begin by formalizing what we mean by equivariance with respect to transformations over time. Succinctly, flow equivariance extends the existing `static' group equivariance to time-parameterized sequence transformations (`flows'), such as visual motion. These flows are generated by vector fields $\nu$ which define the evolution of an autonomous ordinary differential equation of $\mathbf{x} \in \mathbb{R}^n$: 
\begin{equation}
\frac{d}{dt} \mathbf{x} = \nu(\mathbf{x}), \qquad \mathbf{x}(0) = \mathbf{x}_0
\end{equation}
We write the operator associated with this vector field $\nu$ as $\psi_t^{(\nu)}$ (often dropping the dependence on $\nu$ for simplicity), which can be simply thought of as mapping from some initial state $\mathbf{x}_0$ to the state at time $t$ after undergoing the flow for $t$ timesteps (i.e. $\psi_t(\mathbf{x}_0) = \mathbf{x}_t$). A flow then acts on signals $\mathbf{u}(\mathbf{x})$ defined over the space $X$ via the pullback action $(\psi_t \cdot \mathbf{u})(\mathbf{x}) = \mathbf{u}(\psi_{-t}(\mathbf{x}))$. In other words, the flow can be seen to transform the coordinates over which the signal is defined, causing it to `move' (Box \ref{fig:flow_vis}).

A sequence-to-sequence model $\Phi$ such as a Recurrent Neural Network (RNN), mapping from an input sequence $\{\mathbf{u}_t\}_{t=0}^T$ to set of hidden states $\{\mathbf{h}_t\}_{t=0}^T$, is then said to be flow equivariant if, when the input sequence undergoes a flow, i.e. $\{\mathbf{u}_t\}_{t=0}^T \rightarrow \{\psi_t \cdot \mathbf{u}_t\}_{t=0}^T$, the hidden state sequence also transforms according to the action of a flow: $\{\mathbf{h}_t\}_{t=0}^T \rightarrow \{\psi_t \cdot \mathbf{h}_t\}_{t=0}^T$. We can write an equivariance relation, as before, giving us: 
\begin{equation}    
\Phi\bigl[\{\psi_t \cdot \mathbf{u}_t\}_{t=0}^T\bigr]_T = \psi_T \cdot\Phi\bigl[\{\mathbf{u}_t\}_{t=0}^T\bigr]_T \quad \forall\  T
\end{equation}

\begin{figure}[h!]
\fbox{%
\begin{minipage}{0.99\textwidth}
\begin{center}
\textbf{\color{brown} Box 4. Flow Equivariant Recurrent Neural Networks} \vspace{-3mm} 
\end{center}
\hrule 
\vspace{1mm}
\hspace{0.1cm}
\includegraphics[width=0.98\textwidth]{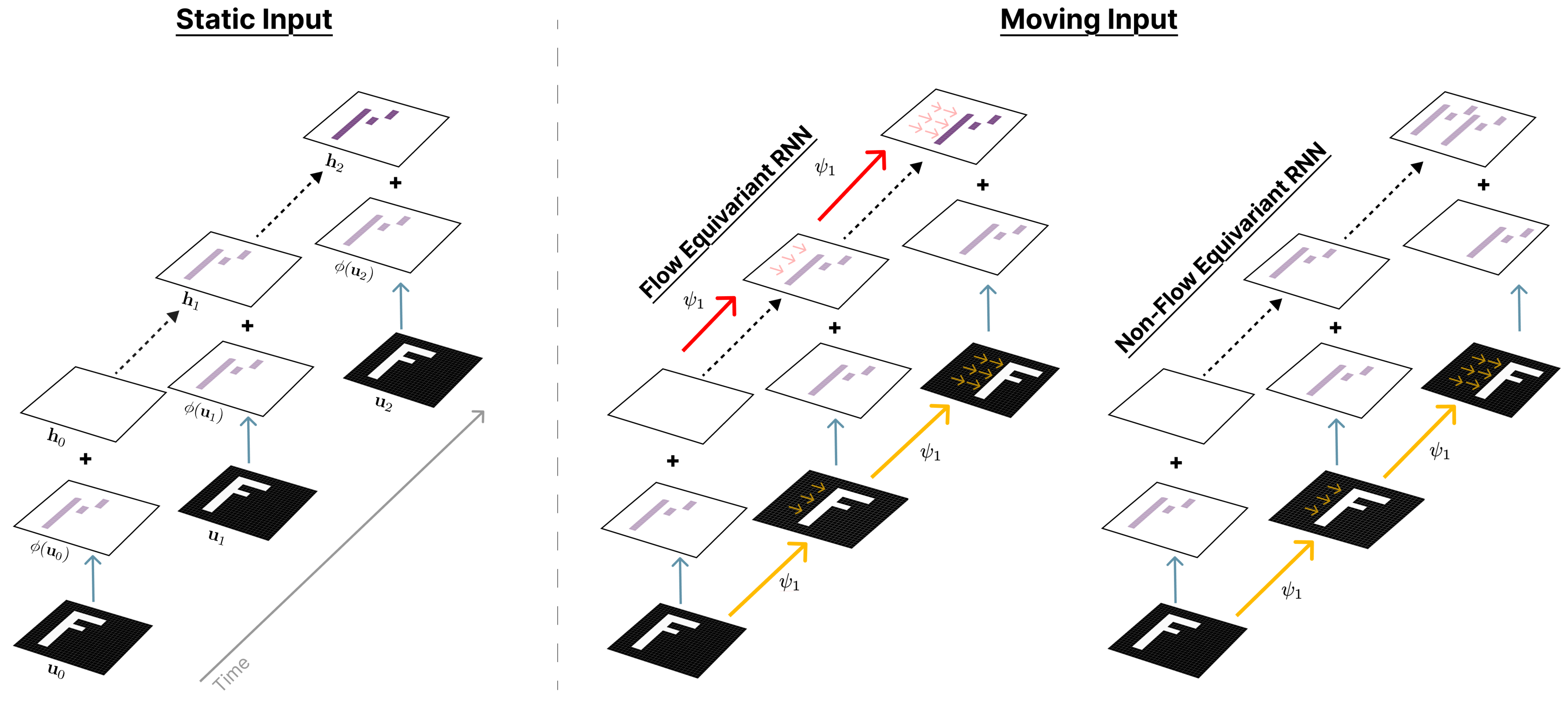}
\vspace{-0.3cm}
\captionsetup{labelformat=empty}
\caption{\textbf{A flow equivariant neural network will process a moving stimulus sequence the same way as it processes a static stimulus sequence, but with the same motion applied to its hidden state. To achieve this property for a specific input flow $\psi^{(\nu)}$, a corresponding flow must preemptively shift the hidden state before the input comes in (denoted by red arrows).} We show (left) the hidden state of a simple RNN $(\mathbf{h}_{t+1} = \mathbf{h}_t + \phi(\mathbf{u}_t))$ processing a static image. For flow equivariance to hold, the hidden state of the network processing a moving image should be a moving version of this output. 
We see (middle) that a flow equivariant RNN $(\mathbf{h}_{t+1} = \psi_1 \cdot \mathbf{h}_t + \phi(\mathbf{u}_t))$  does indeed achieve this property -- the hidden state flows in unison with the input flow (denoted by the red arrows in the feature space). On the right, we see that a regular simple RNN does not satisfy this property, since the hidden state at each timestep appears to `lag behind' the moving input, leading to a blurring of the hidden state where the input and the hidden state do not add constructively as they did for the static case.
}
\label{fig:flow_equivariance}
\end{minipage}%
}
\end{figure}

We see this is the direct analog of the equivariance relation stated earlier, simply extended to models that process sequences -- if the input undergoes a transformation, there is a corresponding `equivalent' (homomorphic) transformation of the output. We again note $\psi$ may act differently on $\mathbf{u}$ and $\mathbf{h}$ but we use the same action symbol for convenience. We explicitly visualize this in Box \ref{fig:flow_equivariance}. %

As can be seen by the diagram, to achieve flow equivariance, it turns out to be necessary to perform computation in the co-moving reference frame of the input. In other words, for a simple RNN, the hidden state must flow in unison with the input (Box \ref{fig:flow_equivariance} middle). For an input flow $\psi_t$, a flow equivariant simple RNN must look something like:
\begin{equation}
\label{eqn:fernn}
    \setlength{\abovedisplayskip}{2pt}
    \setlength{\belowdisplayskip}{2pt}
    \mathbf{h}_{t+1} = \sigma \bigl(\mathbf{W}(\psi_{1}^{(\nu)} \cdot \mathbf{h}_t) + \mathbf{Vu}_t\bigr),
\end{equation}
where $\mathbf{W}$ is a recurrent connectivity matrix, and $\mathbf{V}$ is a matrix mapping from the input to the hidden state. More generally, for an arbitrary recurrent update function $F(\mathbf{h}_t; \  \mathbf{u}_t)$, the update must satisfy: $\mathbf{h}_{t+1} = F\bigl(\psi_{1}^{(\nu)} \cdot \mathbf{h}_t; \  \mathbf{u}_t\bigr)$.

What this means, is that in the simplest possible linear RNN setting with identity matrices (as in Box \ref{fig:flow_equivariance}), the hidden state at the next timestep must be a `moved' version of the hidden state at the previous timestep plus the new input: $\mathbf{h}_{t+1} = \psi_1 \cdot \mathbf{h}_t + \mathbf{u}_t$. Stated explicitly, this concretely shows that in order for a non-trivial recurrent neural network to be equivariant with respect to flows in the input space, \emph{its hidden state must be actively `flowed' through recurrent connectivity in unison with the input flow.} 

We note that the origins of this hidden state flow are unspecified: it may be built-in through connectivity, or inferred from inputs and dynamically modulated implicitly or explicitly. In the case that it is encoded in fixed connectivity, the parameter $\nu$ is likely required to be fixed throughout time. In the case when the hidden state flow is dynamically modulated, the flow may vary over time (denoted $\nu_t$), and we thus may rewrite the above with $\psi_{1}^{(\nu_t)}$, without changing the underlying mathematics. Regardless of the mechanism, the theory stands that some latent flow in recurrent dynamics must exist matching the input flow. We encourage readers to check the formal proofs of the original flow equivariance work\cite{fernn} verifying that these latent flows are indeed not just sufficient but also necessary to achieve flow equivariance in any model that retains a `hidden state', such as canonical artificial and biological recurrent neural networks.

\subsubsection*{Direction Selectivity as Flow Equivariance}
To connect the above requirement to familiar visual computations, consider a 1D retinotopic coordinate
$x$ and a stimulus feature pattern $u_0(x)$ moving at constant velocity $v$ along this one dimension. The resulting input stream
can be written as the action of a translation flow $\psi^{(v)}_t(x)=x+vt$ on the signal via the pullback:
\begin{equation}
(\psi^{(v)}_t \cdot u_0)(x) = u_0(\psi^{(v)}_{-t}(x)) = u_0(x-vt).
\end{equation}
Flow equivariance asserts that a recurrent representation $h_t(x)$ should evolve in a manner homomorphic
to this motion: when the input is ``moved'' by $\psi^{(v)}_t$, the hidden state should undergo the same
transport in its own coordinates (Box \ref{fig:flow_equivariance}). In the simplest additive case this means
\begin{equation}
h_{t+\Delta t}(x) \;=\; \psi^{(v)}_{\Delta t}\cdot h_t(x) + u_{t}(x)
\;=\; h_t(x-v\Delta t) + u_{t}(x),
\label{eq:transport_update}
\end{equation}
i.e. the recurrent dynamics must implement a \emph{co-moving shift} of the hidden state prior
to incorporating new input.
Discretizing space into positions $i$ with spacing $\Delta x$, the pullback becomes a simple index shift by
$k \approx v\frac{\Delta t}{\Delta x}$:
\begin{equation}
h_{t+1}[i] = h_t[i-k] + u_{t}[i].
\end{equation}
This ``shift-by-$k$'' operator is exactly what classic delay-line motion circuitry implements: a lateral
projection from location $i-k$ to $i$ with conduction delay $\tau=\Delta t$ aligns activity so that signals
arrive \emph{in register} only for the matching velocity $v=k\Delta x/\tau$. When such aligned inputs are
combined in a coincidence or correlator nonlinearity, the resulting unit becomes selective to the
\emph{sign} of $v$ (direction), yielding a Reichardt-type direction detector; spatiotemporal energy models
realize the same principle through filters oriented in space-time, which respond maximally to patterns of the
form $f(x-vt)$ \cite{AdelsonBergen1985MotionEnergy}. At the population level, repeatedly applying the same
transport operator across a retinotopic sheet produces a propagating pattern of activity -- a traveling wave.
Thus, for fixed-velocity motion, canonical direction-selective computations can be viewed as
concrete instances of flow equivariance: the circuit implements the co-moving update required
to keep internal representations aligned with a moving input.

With this concrete example in mind, 
it is natural to assert that in the more general case of motion, beyond simple direction selectivity, some spatiotemporal dynamics in neural systems may fill the role of the recurrent latent flows that flow equivariance requires. %
In the following section, we will outline evidence which further supports this perspective. Specifically, we will argue that given the biological constraints on natural neural networks, smooth local spatiotemporal dynamics such as those seen in the brain are the most logical implementation of the necessary latent flow operator. This will naturally lead into our conclusion that the natural bias of large scale neural networks towards structured spatiotemporal activity can be better understood as a natural inductive bias towards a form of flow equivariance. Finally, in the final section, we will show how this spatiotemporal perspective on neural information processing systems presents a unifying perspective -- relating connectivity to selectivity, and spatiotemporal dynamics to memory, in both natural and artificial systems.

%% file: sections/4_The_New_Paradigm.tex
\section*{Spatiotemporal Neural Dynamics are Natural Latent Flows}

In the preceding formal discussion of flow equivariance, no assumptions were made as to the precise form of the necessary flow action in the hidden state. Analogous to existing group equivariant neural networks, we know that there are many valid representations of the group action in the latent space of neural networks as long as they satisfy the group axioms (e.g. valid choices include regular representations\cite{pmlr-v32-cohen14}, or steerable representations\cite{cohen2016steerablecnns, cohen_theory_gcnn}). In this section, we will argue that spatiotemporal flows of neural activity are the most plausible natural candidate to implement latent flows due to both biological constraints 
which place a pressure on spatiotemporal locality, and the spacetime inseparable nature of motion transformations which matches the inseparable nature of flows.

\subsubsection*{Spacetime Inseparability -- The Signature of Motion}
\begin{figure}[h!]
\fbox{%
\begin{minipage}{0.99\textwidth}
\centering
\vspace{1mm}
\textbf{\color{brown} Box 5. Spacetime Separability 
}
\vspace{1.5mm}
\hrule
\vspace{1.5mm}
\includegraphics[width=0.98\textwidth]{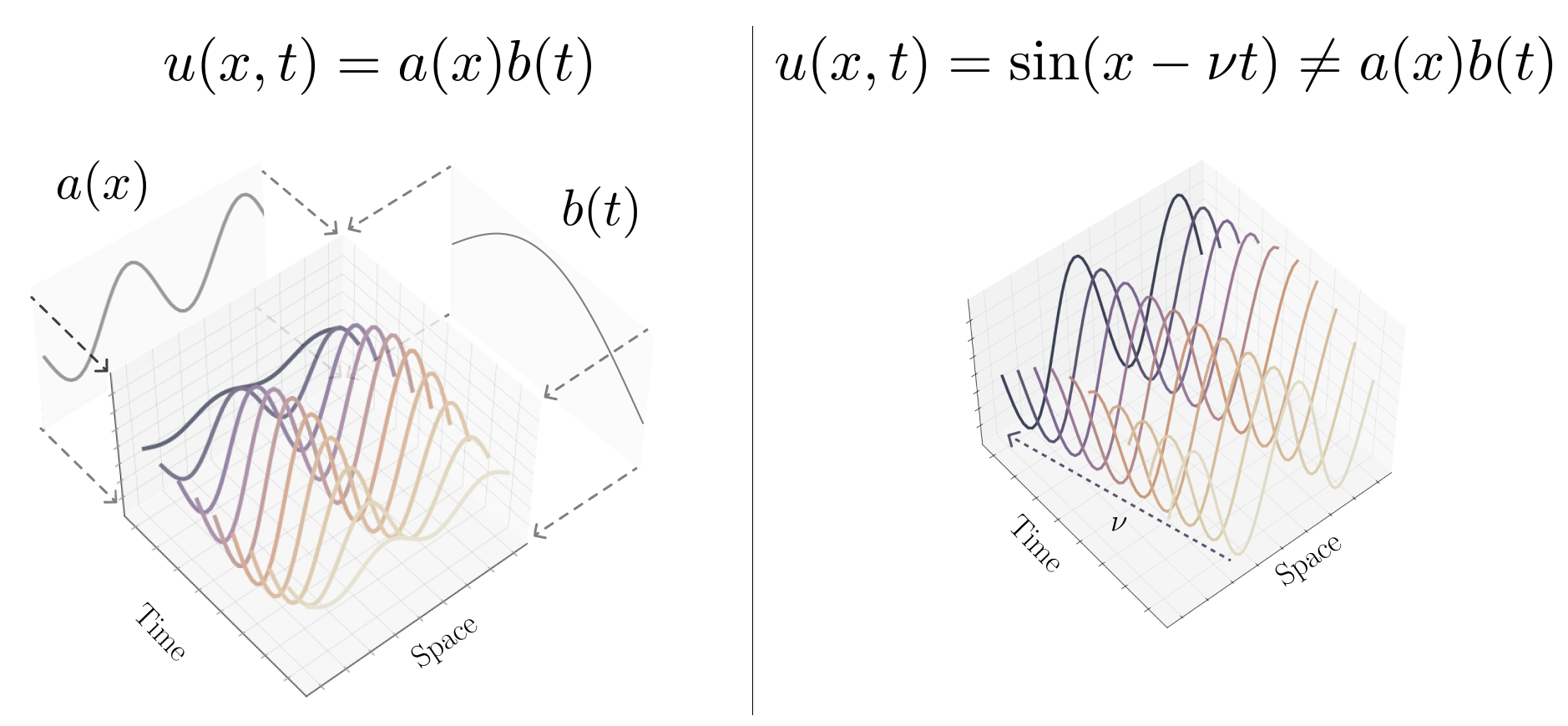}
\captionsetup{labelformat=empty}
\caption{\textbf{Illustration of the fundamental mathematical difference between standing waves and traveling waves, which are not `spacetime-separable.'} For traveling waves, the function $u(x, t)$ that represents the displacement of the field at a spatial location $x$, and individual point in time $t$, cannot be decomposed into two independent functions of space and time; that is, there do not exist functions $a(x)$ and $b(t)$ such that $u(x, t) =  a(x)b(t)$. In particular, the first-order wave equation in the right panel $\frac{\partial u}{\partial t} = \nu \frac{\partial u}{\partial x}$ admits the non-separable general solution: $u(x, t) = f(x - \nu t)$. In the panel on the left, standing waves, the stationary counterpart of traveling waves (sometimes called oscillations), are clearly separable into spatial and temporal components: $u(x, t) = \sin(k x)\sin(\omega t)$. Importantly, generic non-trivial motion transformations (`flows') are inherently spacetime \emph{inseparable}, meaning that if a neural network would like to represent them consistently, its dynamics must also be spacetime \emph{inseparable}.
}
\label{fig:spacetime_separable}
\end{minipage}%
}
\end{figure}

At a high level, the defining characteristic of traveling waves in neural systems is that of spacetime \emph{inseparability}, meaning that the function representing neural activity over space and time ($u(\mathbf{x}, t)$) cannot be decomposed into two independent functions of space and time alone, i.e. $u(\mathbf{x}, t) \neq  a(\mathbf{x})b(t)$ (Box \ref{fig:spacetime_separable}). Interestingly, spacetime inseparability is the same quality which defines non-trivial flow transformations of stimulus signals. More specifically, flows explicitly mix space and time coordinates of signals. For example, a simple translation flow $(\psi_t^{(\mathbf{v})}: \mathbf{x} \rightarrow \mathbf{x}-\mathbf{v}t)$ can map a given signal $u(\mathbf{x}, t)$ defined over space, to a signal $u_{\psi}(\mathbf{x}, t) := u(\mathbf{x} -\mathbf{v}t, t)$ where the spatial coordinate now depends on time. The resulting signal $u_{\psi}$ is thus generally only spacetime-separable if the translation velocity $\mathbf{v}$ is $\mathbf{0}$. For general flowing signals, it can similarly be shown that they are only spacetime-separable when $\psi_t$ is constant for all $t$ -- in other words, when $\psi$ is a static symmetry. 
There exists exceptional solutions where the initial condition $u(\mathbf{x}, 0)$ is an eigenfunction of the dynamics (e.g. $u(\mathbf{x})$ is constant over space in the translation example), but generally, a spacetime-separable signal cannot be a consistent representation of a generic flow transformation. 
This simple realization leads to a relatively significant restriction of the types of neural dynamics which may be able to consistently represent non-trivial flows of the input. For example, save limited exceptional settings, traditional oscillatory `standing wave' activity would not be able to consistently represent generic motion of an input signal. 

There is additionally strong evidence that a large portion of neural activity is indeed spacetime inseparable, and that respecting this inseparability during analysis is necessary in order to derive correct conclusions. For example,  a study by Makeig et al. (2002)\cite{doi:10.1126/science.1066168} demonstrated a profound difference in the interpretation of the source of event-related potentials (ERPs) when analyzing trial-averaged EEG responses in comparison with the single-trial analysis of the recordings using independent component analysis (ICA). ICA revealed phase resetting of an underlying alpha rhythm by a visual stimulus; that is, the stimulus did not elicit the peaks, but shifted the ongoing activity. ERP peaks are typically separated by 100 ms, the period of the 10 Hz alpha rhythm, suggesting that these peaks may reflect phase resetting of the ongoing alpha traveling waves rather than distinct evoked responses. Alexander et al. (2015)\cite{donders_is_dead} later highlighted the consequences of assuming spacetime separability with techniques such as trial averaging, emphasizing that when the separability assumption is violated, these methodologies not only introduce additional noise to measurements, but can also lead to false conclusions. A concrete example of this can be seen in Muller et al. (2014) \cite{Muller2014}, where wave dynamics are invisible in the trial-averaged responses, but stand out at the single trial level.

While this is strong evidence that spatiotemporal neural dynamics may be a natural choice for representing moving flows of input stimuli, it does not place any constraints on the spatial layout of the dynamics themselves. In other words, a spatially unstructured set of neural dynamics could be used to represent a movement transformation as long as in some basis it is a valid flow transformation. Indeed, recent modeling work has trained simple artificial RNN models without any biological constraints, and found latent traveling-wave-like flows to propagate in an alternative `hidden' basis\cite{karuvally2024hidden}. In the following, we argue that due to metabolic efficiency arguments, similar to those believed to underlie topographic organization \cite{KOULAKOV2001519, CHKLOVSKII2002341}, there will also be a pressure towards spatiotemporal locality of neural dynamics such that the putative spacetime inseparable latent flow will thus also be constrained to be a smooth spatiotemporal flow, i.e. a traveling wave that roughly propagates along the cortical surface.

\subsubsection*{Metabolic Efficiency and Time Delays Encourage Spatiotemporal Locality}

One of the most established constraints of biological systems is the pressure to achieve metabolic efficiency. In the context of neural systems, this constraint has engendered highly impactful ideas such as the efficient coding hypothesis \cite{Barlow1961PossiblePrinciples} describing the emergence of simple cell receptive fields \cite{OlshausenField1996SparseCodeNature}, and the hypothesis of wiring length minimization \cite{KOULAKOV2001519} describing the emergence of topographic organization of cortical selectivity. In the context of topographic organization, one of the core principles is that long-range connections between neurons are more metabolically costly to maintain than shorter range connections, and therefore neurons which must communicate frequently with one another should be spatially located close to one another in order to maximize efficiency. We argue that this same principle can be used to explain the existence of a pressure towards spatiotemporal locality of neural dynamics. Specifically, if two neurons must fire frequently in sequence, and neural activity must preemptively flow between them due to flow equivariance, then it is more metabolically efficient for these neurons to be located within close spatial proximity to one another. Indeed, recent modeling work has reinforced this hypothesis, demonstrating that traveling-wave-like dynamics emerge in recurrent neural networks trained to perform a memory task only when a `wiring length cost' term is added to the optimization objective \cite{Shervani-Tabar2026.01.08.698281}.

Further, beyond the direct maintenance of long range connections, it is reasonable to assume that the additional biological constraint of distance-dependent time delays plays a reinforcing role. Since spikes have a finite axonal conduction velocity, if signals need to reach their target within a particular time delay, then this may require myelination of lateral connectivity. This myelination comes at an additional metabolic cost, and therefore undoubtedly yields an additional pressure to reduce the length of lateral connections. 
While speculative, this hypothesis constrains the set of possible implementations of latent flows to match the spatiotemporally coherent traveling waves often observed in neural systems.

%% file: sections/Mechanistic_substrate.tex
\section*{Spatiotemporal Dynamics as an Inductive Bias of the Biological Substrate}
In the above, we have argued that latent flows of neural activity are necessary for the accurate modeling of stimuli undergoing motion transformations, and further, because of biological constraints, these latent flows are most naturally implemented by the coherent spatiotemporal structures in neural activity observed in the neuroscience literature, e.g. traveling waves. From this perspective, it is natural to wonder if traveling waves have arisen as an adaptive solution to achieve flow equivariance, or if instead they are natural to the biological substrate, and thus co-opted to serve as a natural inductive bias towards flow equivariant modeling when beneficial. In this section we will provide evidence for the latter, suggesting that while both are likely true to some extent, flowing spatiotemporal dynamics do seem inherent to large scale biophysical neural network dynamics, and thus it is likely that such an inductive bias would play a significant role in shaping the learned representations of these systems.

\subsubsection*{Time Delays Greater than Membrane Time Constants Yield Spatiotemporal Dynamics}

\begin{figure}[h!]
\fbox{%
\begin{minipage}{0.99\textwidth}
\centering
\vspace{1mm}
\textbf{\color{brown} Box 6. Time Delays in Neural Systems 
}
\vspace{1.5mm}
\hrule
\vspace{1.5mm}
\includegraphics[width=0.65\textwidth]{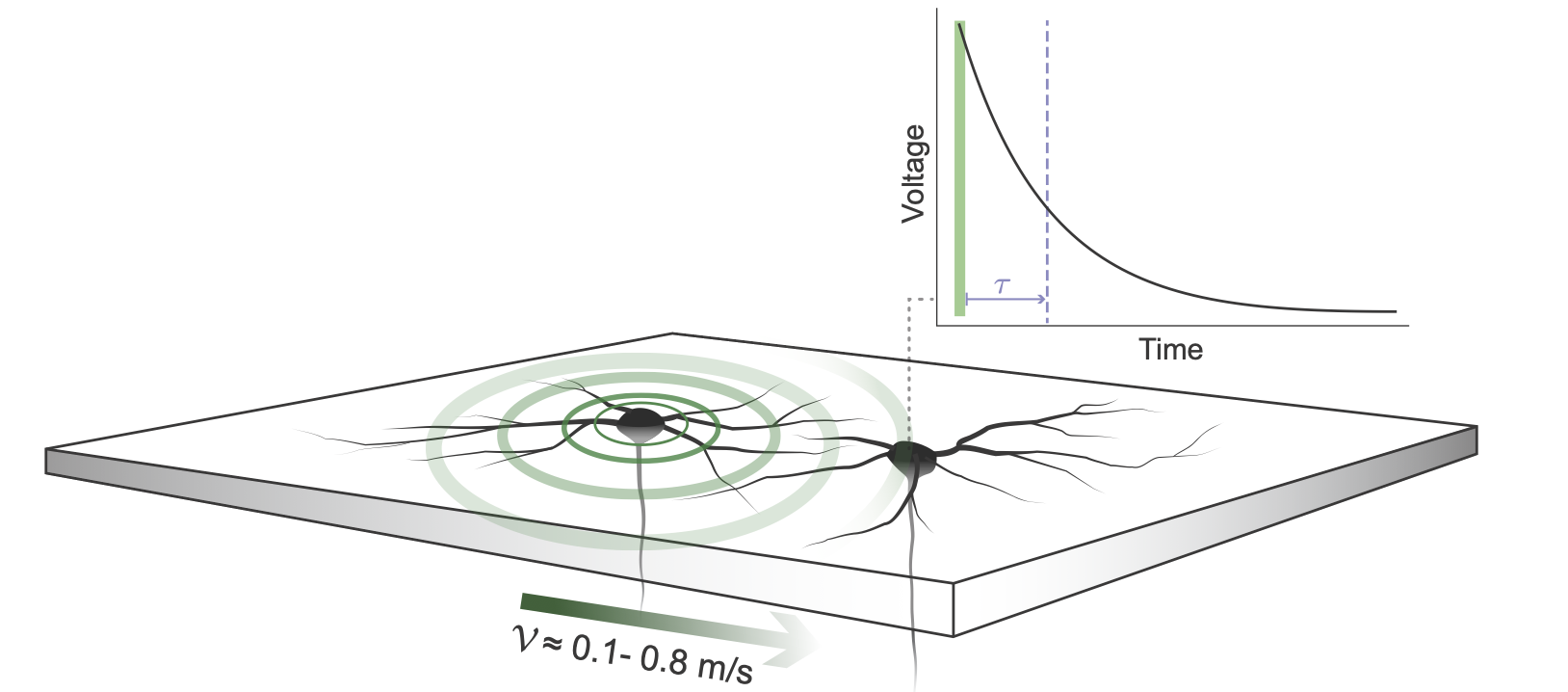}
\captionsetup{labelformat=empty}
\caption{
\textbf{Biological neural systems have inherent time delays that constrain control and communication.} At the most fundamental level, any physical system in which time-delays are non-negligible, i.e. large relative to the characteristic timescale of the dynamics, will exhibit spatiotemporal structure\cite{ROXIN2011323, Campbell2007, Bressloff_2012}. In the brain, this is dependent on the interplay of conduction velocity and the neural membrane time constant. %
Specifically, the membrane time constant, the characteristic time of the exponential decay of a neuron back to its resting state potential, is on the order of 10 ms\cite{Eyal2016UniqueMembraneProperties}, while local traveling waves have been measured to propagate at roughly 0.1 to 0.8 m/s \cite{Muller2018}. The spatiotemporal organization of neural activity into waves then makes sense as a result of this timescale imbalance and the distances signals must travel. 
}%
\label{fig:time_delays}%
\end{minipage}%
} 
\end{figure}

At a high level, we know that the speed of neural transmission $v$ along unmyelinated lateral connections in the cortex is slow relative to the membrane time constant $\tau$\cite{girard2001feedforward, Liewald2014} (Box \ref{fig:time_delays}). 
In humans, horizontal `patchy' connections within individual cortical regions, such as the primary visual cortex, can extend over lengths $L$ which exceed the distance that can be traveled in the time given by the membrane time constant, in other words, $L > v \cdot \tau$. As transmission delays increase relative to such intrinsic cellular time scales, recurrent neural systems can enter a regime of delay-induced oscillatory dynamics (Hopf bifurcations)\cite{Campbell2007}, and in spatially extended neural field models with distance-dependent axonal delays this can manifest as spatiotemporal patterning such as traveling waves\cite{Bressloff_2012}.
If the speed of transmission were significantly faster, the delay in spiking activity induced by spatial separation would be negligible compared with an individual neuron's response time, and spatiotemporal dynamics would likely not be prominent. 

It is natural to question then why these pathways are not biologically modified to avoid delays.  
To understand this, consider there are two known primary mechanisms to increase the relative speed of information transfer in neural circuits: increasing the diameter of axons which transmit signals, and myelination, where auxiliary cells wrap axons with a lipid sheath to increase the conduction speed of spikes \cite{waxman1980determinants}. It is known that conduction velocity is linearly related to axon diameter for myelinated neurons\cite{nowak1997timing}, but for unmyelinated neurons, conduction velocity increases sub-linearly, varying with the square root of the axon diameter \cite{rushton1951}. Myelination can increase transmission speeds up to 150 m/s \cite{Purves2001}; however, this also comes at a cost of increasing the diameter of axonal connections \cite{budd2010}. Following these observations, it can be argued that in order to make all neural transmission time delays negligible, the volume requirements of these connections would be so significant that the resulting proportion of brain volume made solely of axons would be highly suboptimal for natural behavior. Overall brain volume would be excessively large to maintain an equivalent number of neurons, and more generally make natural behavior impossible\cite{CHKLOVSKII2002341}. Further, others have argued that this tradeoff between brain volume, interconnectedness, and time delays may be a reason for increased specialization in larger brains\cite{ringo}.

In this sense then, transmission delays which are significantly greater than membrane time constants appear to be a natural consequence of biological neural systems, and traveling waves would be expected to be measured along these axons with long time delays. Indeed, it has been measured that local (mesoscopic) traveling waves in the cortex propagate at roughly 0.1 to 0.8 m/s, in line with the conduction velocities of unmyelinated lateral connections\cite{girard2001feedforward, Muller2018}; and corroborating evidence demonstrates conduction delays of horizontal fibers in the superficial layers of the cortex are the primary factors which control the generation of wave-like dynamics in the brain \cite{Muller2014, doi:10.1126/science.283.5402.695}.

\subsubsection*{Waves as Natural Solutions to Large Neural Network Dynamics}
In the decades during which the initial interest in spatiotemporal neural dynamics was overshadowed by single neuron electrophysiology, theoretical neuroscientists were unhindered by the technological limitations of recording technology, and therefore continued to study the spatiotemporal solutions to large scale abstract neural dynamics. For example, as early as 1972, Wilson and Cowan wrote down what they called `Neuronal Tissue Equations', modeling the cortex as a one or two-dimensional homogeneous sheet with lateral connectivity \cite{wilson1972excitatory}. One year later, the same authors then demonstrated analytically that when such a sheet is provided with a localized stimulus and sufficiently strong global inhibition, traveling waves exist as solutions to these equations, propagating outwards from the stimulus location without attenuation, such that their frequency and wavelength are non-linear functions of the stimulus impulse itself, rather than connectivity \cite{wilson1973}.
Shortly thereafter, Ellias and Grossberg (1975)\cite{ellias_grossberg} studied similar equations and observed "[p]ersistent travelling waves were generated, and their form depended on the initial data. [...] These reverberations indeed `remember' their initial data, in the sense that they generate different wave structures." Later work from Amari (1977) \cite{amari1977}, Ermentrout (1998)\cite{Ermentrout_1998}, Roxin et al. (2005)\cite{roxin2005role}, and others further reinforced these findings in a variety of different network equations. 
Thus, even from these early examples, it has been clear that traveling waves and structured spatiotemporal dynamics are realistic, often common, solutions to biophysical models of neural systems. %

More recently however, computational work has begun to demonstrate that these types of dynamics arise not only as solutions to approximate mean-field equations, but also emerge as the result of simulations of large numbers of biophysical spiking neurons. For example, Davis et al. (2021)\cite{Davis2021} studied a large-scale Leaky Integrate and Fire (LIF) spiking neural network model and demonstrated that, with topographically organized connectivity and distance-dependent time delays, networks at the biologically realistic scale naturally produce traveling waves statistically similar to those observed in cortex (Box \ref{fig:recorded_waves}, bottom). Notably, the spontaneous waves produced by this model precisely matched cortical waves in terms of their sparsity; specifically, in awake, behaving animals, there is a relatively low probability that any given neuron will participate in a wave. This is in stark contrast to the dense waves observed, for example, in epileptic conditions\cite{pinto2005initiation}.

\subsubsection*{The Inductive Bias of Spatiotemporal Structure}

Taken together, the above mathematical, computational, and biological reasons strongly support the idea that spatiotemporal structures such as traveling waves are inherent to neural dynamics. Given this, we find that the best way to understand the potential `role' of such dynamics may be as an inductive bias -- a set of constraints imposed on a learning system a priori which define its generalization and data efficiency properties~\cite{wolpert1996}. It is thus natural to ask: if a learning system is inherently biased towards the production of spatiotemporal dynamics, how might the learned representations be shaped or constrained? %

Returning to our definition of flow equivariance, we see that a flow equivariant recurrent neural network requires a corresponding representation of the input flow in the output space. From the preceding text, we have repeatedly argued that spatiotemporal dynamics are likely to serve precisely this role: we have shown that their spacetime inseparability makes them the ideal for representing input flows, and we have argued that they are natural to large scale neural network dynamics. But how would it come to be that intrinsic flows of a neural network's activity would end up learning to represent time-parameterized symmetries?

To understand this, we must consider the influence of time, and specifically how an input stimulus (which the neural system is purportedly attempting to represent) changes over time. If we, like J. J. Gibson \cite{gibson1979ecological}, consider an ecological approach to vision, the natural transformations of sensory inputs over time are caused by movement, e.g. by self-motion or by external motion of the world. An important aspect of the world, irrespective of our senses, is that it does not tend to change very quickly (a founding principle of Slow Feature Analysis\cite{sfa}). For example, if a child is playing in the garden now, it is very likely she is still playing in the garden a second later, despite the fact the input to my senses might have changed dramatically. We can therefore interpret fast external changes, and also changes caused by our body movements, as approximate symmetries of the true world state, and we may want to build representations which are structured with respect to these generalized symmetry transformations.

\begin{figure}[h!]
\fbox{%
\begin{minipage}{0.99\textwidth}
\begin{center}
\textbf{\color{brown} Box 7. Spatiotemporal Dynamics and Learned Flows} 
\vspace{-3mm} 
\end{center}
\hrule
\vspace{1mm}
\centering
\includegraphics[width=0.95\textwidth]{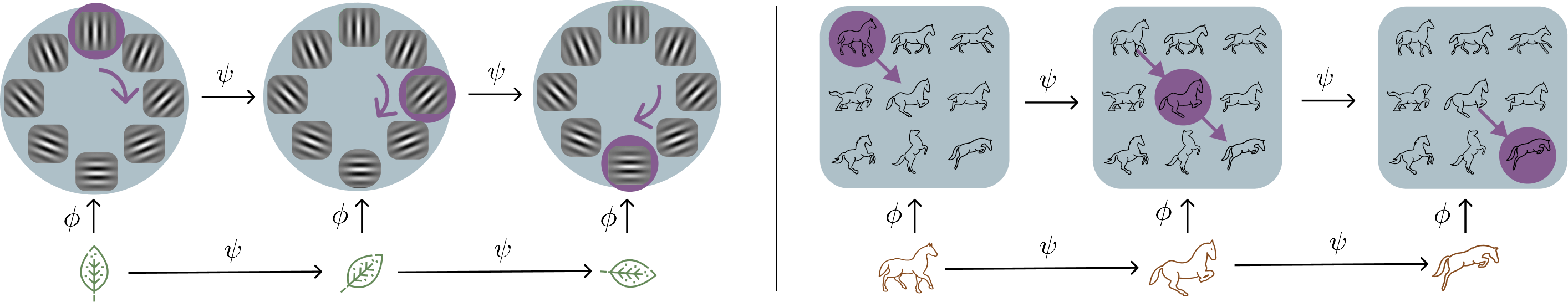}
\vspace{-2mm}
\captionsetup{labelformat=empty}
\caption{\textbf{A visualization of input symmetry transformations and the corresponding spatiotemporal dynamics in a spatially organized equivariant feature space.}  By arranging features according to symmetry transformations, the input can be represented by a smooth shift of activity in the spatially organized feature space. In the simplest case of rotation (left), we see that the corresponding representation will flow as a circular wave between orientation selective cells. For more complex abstract symmetry transformations, such as the changing of the pose of a horse as it prepares for a jump, we see that activity may still flow locally and smoothly in space when organized properly. Importantly, it is the local connectivity, laid out along the cortical surface, that induces this local shift operation, leading to structured representation of the abstract symmetries. We highlight that this is a highly idealized abstraction, and that not all transformations can be laid out perfectly in a 2D grid, and that a highly intricate optimization problem must be solved to figure out the ideal `organization' and feature basis with which inputs are represented; yet, the idea that spatial organization of selectivity is related to lateral connectivity and structured by spatiotemporal dynamics has been supported by both models \cite{nwm} and theory \cite{zucker}. 
}
\label{fig:Equivariance_Waves}
\end{minipage}%
}
\end{figure}

Therefore, our argument concludes as follows: if we consider an agent acting in the world, movement will induce time-parameterized `flow' symmetry transformations of stimuli, and the natural inductive bias of spatiotemporally structured dynamics (traveling waves) will fill the role of the corresponding latent operator 
$\psi$. A learning procedure such as predictive coding which encourages the visual system ($\phi$) to satisfy the commutative diagram (Box \ref{fig:Equivariance_Waves}) will then encourage the visual system to learn a latent structure to satisfy exactly this general form of flow equivariance that we desire. Ultimately, we emphasize that the concept is much more general than the simple translation and rotation symmetries: any smooth `flow' transformation of the input may be mapped to spatiotemporal dynamics in such an architecture.

In this perspective, we put forth the idea that perhaps the observed convergence of structure and function is not coincidental but instead the result of a spatiotemporal inductive bias. More explicitly, if a neural system is biased towards producing structured spatiotemporal dynamics while its driving stimuli undergo approximate flow symmetry transformations simultaneously, we believe this type of flow equivariant latent structure is a natural result. Furthermore, due to the aforementioned known benefits of this type of representational structure in artificial neural networks, we see no reason why this would not similarly benefit natural neural systems to enhance their abstraction and generalization capabilities. 

We believe that the spacetime inseparability of spatiotemporal structured dynamics, and the associated inseparability of the symmetry transformations they could represent, is one aspect which makes this connection so appealing -- if a system has natural latent dynamics which are uniquely homomorphic to a broad class of observed dynamics in the world, it would make sense to leverage this structure as an encoding mechanism. Empirically, a suite of computational work has further provided significant evidence which supports this idea. For example, researchers have shown that by simply imposing strong inductive biases on latent dynamics, without specifying what those latent dynamics should correspond to in the input (i.e. the encoder is randomly initialized), the network learns to encode similar `matching' transformations from the input (e.g. rotations and color transformations) into these structured latent flows \cite{wrnn, keller2021topographic, ffrl, song2025unsupervisedrepresentationlearningsparse}. 
While there are others who have proposed that the orientation column structure and spatiotemporal receptive fields of neurons may support equivariance to small visual transformations in lower level visual areas \cite{covariance}, this idea has not been shown to extend throughout the visual hierarchy, which we believe our perspective supports.

In this article we argue that the benefits of this spatiotemporal structure may extend far beyond the simple visual transformations we have discussed so far. In the next section, we will demonstrate how when memory tasks are considered as flows over \emph{partially observed spacetime signals}  
a set of well known models and empirical results re-emerge as instantiations of flow equivariance.

%% file: sections/Flow_Memory.tex
\section*{Flow Equivariance as Memory}

In this perspective we have described flows as necessary for representing `motion' of input signals, and although we have alluded to this `motion' being general and abstract, we have so far mainly confined our examples and discussion to the visual domain for conceptual clarity. In this section, we will provide concrete examples of more abstract forms of motion which our perspective supports, illustrating the full generality of our abstraction. 

Specifically, we will talk about flows operating on an unobserved `World State' which an agent only partially observes at each point in time. We will demonstrate how by maintaining flow equivariance with respect to potentially unobserved latent flows that act on this world state, but do not show up in observations directly, an agent may build up an accurate, structured, and stable \underline{memory} of the world.

\subsubsection*{Flows in Partially Observable Environments}

A partially observed time-varying signal can be formally described as an unobserved world state $\mathbf{w}_t \in W$ from which observations are produced at each point in time through an observation function $\mathcal{O}(\mathbf{w}_t) : W \rightarrow O$. 
The most familiar example of such a setting is simply our daily experience walking around in the physical world: the world state $\mathbf{w}$ encompasses everything around us, everything behind us, and everything in the next town over. The observation function $\mathcal{O}$ is then simply the projection of that world state onto our senses. %

We argue that although a flow acting on the world state may not be fully represented in our observation (e.g. the motion of someone behind your back has no impact on what you see), there are many cases when it is no less important to be equivariant to such a flow. For example, even in the mundane scenario of crossing the road, while you turn your head to look both ways, it is critically important to be able to maintain an accurate representation of a speeding car that may be coming from behind you. Such a definition of flow equivariance can be formalized as a simple extension to our previous definition earlier in the perspective. Explicitly, a system operating in a partially observed environment is said to be flow equivariant if: when the unobserved world-state sequence undergoes a flow, i.e. $\{\mathbf{w}_t\}_{t=0}^T \rightarrow \{\psi_t(\mathbf{w}_t)\}_{t=0}^T$, the hidden state sequence also transforms according to the action of a flow $\{\mathbf{h}_t\}_{t=0}^T \rightarrow \{\psi_t (\mathbf{h}_t)\}_{t=0}^T$, regardless of whether the full flow is observed in the observation sequence (i.e. it may be that  $\mathcal{O}(\psi_t \cdot \mathbf{w}_t) \neq \psi_t \cdot \mathcal{O}(\mathbf{w}_t)$). In the form of an explicit equivariance relation, we would desire the following equality describing the hidden state of a sequence-to-sequence model $\Phi$ at time $T$: $\Phi[\{\mathcal{O}(\psi_t \cdot \mathbf{w}_t)\}_{t=0}^T]_T = \psi_T \cdot\Phi[\{\mathcal{O}(\mathbf{w}_t)\}_{t=0}^T]_T$. In words, this means that the output of a sequence model processing observations from a flowing world should be equivalent to its output for a static world, but with a flow applied to the output.

In such a setting, the instantiation of flow equivariance may appear confusing: a network's activity may appear to be flowing despite no sign of flowing stimuli. However, when the world state is considered carefully, we can see that indeed such flow equivariance with respect to global unobserved dynamics is extremely powerful for operating in partially observed environments and maintaining memory for things that are no longer in the field of view. In the following, we provide a few concrete examples.

\subsubsection*{Equivariance to World Flows as Memory}
In the simplest example, consider a scrolling LED sign, often seen affixed to the top of food trucks in New York City. Such signs may display the food items that a vendor offers, but often the list may be too long to display all at once. For example, the vendor may sell "Pizza, Hot Dogs, Burgers", but at any given time, the screen is limited to only 12 characters, so it may only show "Pizza, Hot D" when you first look, but over time the screen `scrolls' over this list revealing: "za, Hot Dogs".

We can formalize this as a canonical `partially observed' world with an underlying flow. The world state at time $0$ is given by the full sentence $\mathbf{w}_0$ = "Pizza, Hot Dogs, Burgers" and the observation function $\mathcal{O}$ simply returns a fixed subset of the first 12 characters: $\mathcal{O}(\mathbf{w}_0)$ = $\mathbf{u}_0$ = ["Pizza, Hot D"]. The world state, is then governed by a flow, that shifts the sentence (the signal) left by some number of characters per unit time while the observation window remains fixed. Explicitly: $\mathbf{u}_1 = \mathcal{O}(\psi_{1} \cdot \mathbf{w}_0)$ = "izza, Hot Do", $\mathbf{u}_2 = \mathcal{O}(\psi_{2} \cdot \mathbf{w}_0)$ = "zza, Hot Dog", and so forth.
We can then ask, what value might flow equivariance add in such a setting?  

\begin{figure}[h!]
\fbox{%
\begin{minipage}{0.99\textwidth}
\begin{center}
\textbf{\color{brown} Box 8. Flow Equivariance Supports Structured Memory Formation} 
\vspace{-3mm} 
\end{center}
\hrule 
\vspace{1mm}
\includegraphics[width=0.99\textwidth]{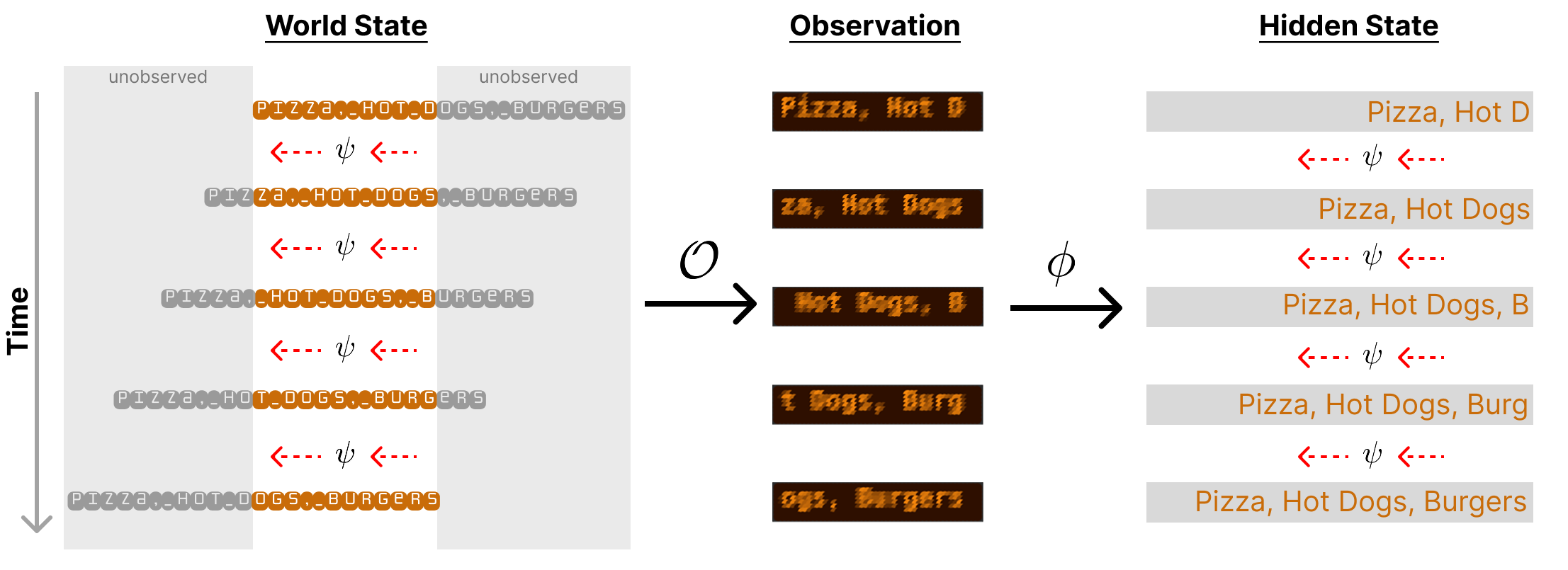}
\vspace{-2mm}
\captionsetup{labelformat=empty}
\caption{\textbf{Flow Equivariance with respect to world state flows can serve as a mechanism for constructing structured memory.} By mirroring potentially unobserved flows of a global world state, a flow equivariant model is able to construct a hidden state that similarly mirrors the true structure of the full unobserved world despite never seeing the entire structure simultaneously. On the left we show the full world state for a simple `scrolling LED sign' -- the sentence: "Pizza, Hot Dogs, Burgers". The sign scrolls through this sentence (through the flow $\psi$, shifting the sentence sequentially to the left over time) and displays only 12 characters at a time from a fixed observation window $\mathcal{O}$. The neural network ingests the observation through the encoder $\phi$, adding to its hidden state (right). We see that when the hidden state mirrors the flow in the input space, the observations are added in the correct order -- maintaining the structure of the input and serving as a structured memory of the past despite not seeing it all simultaneously. Conversely, if the hidden state did not shift in unison with the world state flow, the inputs would overlap or be written to incorrect positions.
}
\label{fig:memory}
\end{minipage}%
}
\end{figure}

First, let us outline what flow equivariance could look like here. One possible instantiation could mean that the network has a large hidden state $\mathbf{h}$, larger than the observation size, which undergoes a homomorphic shift operation `mirroring' the world state shift. We visualize this procedure in Box \ref{fig:memory}. In this way, each new observation will `align' with previous observations, while simultaneously `making room' for new observations. Because of this, even when unobserved, items in the hidden state will still be spatially structured to match the true structure of the world state -- this is the benefit of the homomorphic flow. At the end of the sentence, because of this homomorphism, the full original sentence will be stored in the hidden state in the correct order.

However, we highlight that this is not only relevant for visual memory, this mechanism applies for storing any type of structured memory. For example, if you are trying to remember a phone number that is spoken one digit at a time, it may be natural to see this as a flow operating on the world state of the number, revealing itself to you one digit at a time. To remember the digits in the correct order, you should have a corresponding flow in your hidden state -- effectively `writing' this memory into a spatial pattern of activity, translating experienced time into space. 

More generally, we believe this idea also extends to spatial memory. In the case of a rat navigating a maze, the flow on the world state can be entirely described by the agent's own actions -- if it walks forward, the world equivalently flows backwards relative to the agent's egocentric frame; if it turns right, the world flows exactly inversely left. The agent can therefore use its own motion signals or `efference copies' to infer how the world will flow at each point in time, and thereby apply this estimated flow to its own hidden state in order to maintain flow equivariance.
Doing this properly means that when an agent gets back to where it started in the physical world, its latent map will also return back to where it started. It will have correctly closed the loop and built a structured latent space -- a proper `cognitive map'\cite{Tolman1948CognitiveMaps}.
Finally, we hypothesize that the same idea may apply to logical reasoning: as you reason through a sequence of logical arguments, each step may be seen as taking an action in an abstract space, and this will correspond to a motion or a flow. In order to build such a `logical map' \cite{Bellmund2018NavigatingCognition} that facilitates reasoning in the future, you will thereby want your representation of the argument to flow in unison with these reasoning actions.

In real world scenarios, the ground truth `world state flow' is not always given or even uniquely determined from a given stimulus sequence. For example, in the case of someone speaking their phone number and stating ``1, 1, 1'' it may be ambiguous whether they are advancing to the next digit in the sequence, or whether they are simply repeating the first digit multiple times waiting for a confirmation. It is therefore critical that systems learn to infer the likely world flows underlying their observations, and we believe that in many daily scenarios this is likely possible with high fidelity. However, errors in this approximation would undoubtedly be a source of great confusion, and reducing such error may constitute a large portion of what it means to learn how to reason. The aforementioned `cycle consistency' may be used as one potential signal for determining when a sequence of unobserved latent flows have been estimated correctly, and thereby may serve as a teacher for such estimators. 

Overall, this idea of a spatially structured memory is in line with ideas that have existed in neuroscience for a long time. It is reminiscent of the idea of `memory palaces'\cite{Yates1966ArtOfMemory}, or the `Method of Loci'\cite{methodofloci}, a prominent technique used in memory competitions, where long sequences are recalled by visualizing sequence elements in orderly locations of an imagined spatial trajectory. 
Relatedly, Eichenbaum argued more plainly that ``The role of the hippocampus in navigation is memory'' \cite{eichenbaum}; and others have extended this to generalize `cognitive maps' beyond spatial navigation to non-spatial knowledge \cite{cogmap, doi:10.1126/science.aaf0941, TAVARES2015231}. Recent perspectives argue that representations from grid cells may form a structured `scaffolding' that memories can be pinned to, like clothes pinned to a clothes line \cite{Chandra2025VectorHaSH}; and recent theoretical and computational work has further demonstrated that grid cells form as a consequence of geometric `equivariance' constraints \cite{Dorrell2022ActionableGridCells}. In alignment with these previous viewpoints, we find it very reasonable to believe that evolution may have repurposed systems for navigation to serve as methods for maintaining structured memory\cite{BuzsakiMoser2013Theta}, or more generally structured representation learning\cite{Song2025StructuredRepLearning}. 

Perhaps most relevant, Whittington et al. (2025)\cite{Whittington2025Slots} argue that prefrontal cortex may form a type of slot-based memory such that sequences can be stored in a structured, controllable manner. Their conceptual model can be seen as similar to a special case of the proposed partially observed equivariant flows. Specifically, our formalism generalizes these ideas beyond simple 1 and 2D space to any structured domain, and argues that spatiotemporal flows (such as those observed in prefrontal cortex) are likely the operator which instantiates the permutation or shift required for equivariance. Experimentally, potential links have been made between propagating traveling waves in the prefrontal cortex and working memory task performance in monkeys\cite{distraction_waves, working-memory-waves}, hinting at the likelihood for this more general connection.

Ultimately, while this idea may appear on first glance as an unnecessary application of abstract mathematics to a simple problem, recent work has begun to provide strong evidence for its applicability to understand both natural and artificial systems. In the following section, we highlight key evidence from both neuroscience and artificial intelligence which we believe strongly supports the idea that equivariance to spatiotemporal transformations may be the right lens with which to begin to unravel the mysteries of spatiotemporal dynamics in neural networks.

%% file: sections/6_Unifying_AI_and_Neuro.tex
\section*{A Unifying Perspective}

When we take the above spatiotemporal perspective on neural dynamics and neural representations, we see that it provides a unified view of diverse findings in the fields of both natural and artificial intelligence. As we have already outlined, our perspective makes a link between the observed traveling-wave-like dynamics of biological neural networks and equivariant structure in artificial neural networks. In the following we will describe how both recent and historical findings on dynamics and structured recurrent connectivity may be unified in a spatiotemporal perspective. Finally, we will show how many state-of-the-art artificial models such as State Space Models (SSMs) and Transformers may be understood in terms of spatiotemporal dynamics, providing insights into their connection with the brain and how these models may be extended.

\subsubsection*{Signatures of Flow Equivariance in the Structure of Lateral Connectivity}

One branch of research, often called `neurogeometry' \cite{Petitot2017}, has precisely studied the connectivity of visual cortex in terms of the geometric structure of the environment that it aims to represent. Proponents argue that the early visual cortex can be described most naturally by the language of differential geometry, specifically in terms of fiber bundles and contact structures\cite{hoffman_contact, sarti, 3d_continuation}, where the precise connectivity structure is governed by the `natural' geometric structure of our 3D environment and 2D observations. 
This framework can be understood at a high level to assert that transformations which define perceptual constancies (e.g. shape, scale, motion, color) are `encoded' in the cortex through lateral connections between neurons, thereby serving to facilitate the detection of smooth continuous lines, contours, and shapes, forming the basis of Gestalt `good continuation' and `association fields'\cite{Field1993}. These ideas as a whole can be seen as the beginnings of an understanding of how the brain structurally represents the transformation properties of low level features as the building blocks for high level objects.

In detail, this `contact bundle' description is based on the finding that lateral connections preferentially connect neurons with similar selectivity\cite{bosking1997orientation}.
The indirect functional role of these connections has been measured for a long time in terms of what are called `non-classical receptive fields' or `silent surround receptive fields'\cite{SERIES2003453}, whereby a neuron's activity can be modulated by other specific neurons outside of its primary (classical) spatial receptive field. However, the understanding of the role of these connections to canonical recurrent processing has to date been limited -- similar to the limited understanding of the role of spatiotemporal dynamics broadly. Yet recently, it has been shown that these lateral connections are indeed precisely those which mediate traveling wave dynamics \cite{Muller2018, Muller2014, doi:10.1126/science.283.5402.695}; and further, it has been measured that waves propagate precisely along these preferred directions in orientation space \cite{Davis2024}. Empirically, it has been measured that oriented lines which travel along these preferred directions are indeed perceived to move more quickly due to the potentiating effect of traveling waves \cite{SERIES20022781}; and computationally, models have been built which demonstrate the forward predictive power of lateral connectivity in a manner very similar to this\cite{Benigno2022, nwm}. Finally, most surprisingly, it has been measured that such lateral connections appear to facilitate activity at saccade-like speeds exactly along these co-linear orientation trajectories \cite{Gerard-Mercier3925}. 
So, what does this mean in consideration of our perspective on spatiotemporal dynamics and flow transformations? 

In the language of differential geometry, the well known orientation columns of the early visual pathway can be seen as `lifting' smooth curves in input space into the matching smooth curves within an orientation `fiber bundle' (analogous to the rotation equivariant neural network depicted in Box \ref{fig:commuting_diagram-a}). The lateral connections in V1 then trace out smooth integral curves of a sub-Riemannian manifold in this latent position and orientation space\cite{3d_continuation, sarti}.  Importantly, these smooth curves are defined by simple generating vector fields -- flows, defined now not over time, but over space. This precise `sub-Riemannian' structure of connectivity matches feature co-occurrence statistics from natural images\cite{Sarti2011}, and further, precisely predicts the distribution and arrangement of lateral connectivity in the primary visual cortex\cite{zucker}.

Succinctly, the neurogeometry literature has demonstrated that lateral connectivity in visual cortex is arguably best described in terms of `natural' smooth flow transformations of infinitesimal visual elements, where the parameter $t$ which defines the flow ($\psi_t$) is not time but space. We also know that these same lateral connections likely support traveling waves, providing a strong link between flow transformations of all kinds and spatiotemporal neural dynamics, supported by studies of neural connectivity. We believe that evidence for these `frozen' flows is only the first step, perhaps easiest to discover with existing neural recording methods. With new recording technology yielding improved temporal and spatial resolution, we believe that the precise signatures linking neural dynamics and world flows \emph{over time} will come to light.

In the following, we discuss how spatiotemporal dynamics and the associated benefits are found not only in biological neural systems, but are also increasingly being built implicitly or explicitly in artificial systems.

\subsubsection*{Signatures of Flow Equivariance in Artificial Neural Networks}

A natural place to look for spatiotemporal dynamics in artificial neural networks is in the recurrent dynamics of their hidden activity. If spatiotemporal dynamics were to exist in such models, following the logic of the previous sections, they would likely serve to encode flows, invariants, and symmetries of the task which they are trained to perform; and indeed, numerous recent studies have found this to be the case. 

In the context of visual flow symmetries, work by Keller and Welling (2021)\cite{keller2021topographic} illustrated that by inducing a spatio-temporal prior over neural activity in a generative model (described as shifting-topographic distribution), structured equivariant subspaces (`capsules') were indeed learned to reflect the flow symmetries of their training data. In subsequent work (2023)\cite{nwm}, the same authors demonstrated that by enforcing a local inductive bias on the coupling parameters of an oscillatory recurrent network, and subsequently training the network to accurately predict future states of an ongoing symmetry transformation (effectively `closing the loop' presented in Boxes \ref{fig:commuting_diagram-a} \& \ref{fig:flow_equivariance}), the network learned to propagate traveling waves over the simulated `cortical sheet'. They further demonstrated that these waves did indeed spatially structure the selectivity of neurons in the network to respect the symmetries at hand through (i) visualized selectivity maps reminiscent of orientation columns\cite{keller2023locally}, and (ii) by artificially inducing traveling waves in the hidden state and observing the reconstructed stimuli undergo the same transformations the network was trained on\cite{nwm}. In a similar biologically plausible network implementation of the hippocampal formation, Chen et al. (2024)\cite{chen2024predictive} demonstrated that when a recurrent autoencoder network is trained to predict the next state of an input stream by encoding the input, propagating activity in the hidden states, and then subsequently decoding the end result, the model learned a Toeplitz recurrent connectivity structure as a way for the recurrent hidden layer to store the sequence and later recall it. This Toeplitz structure can be related directly to traveling waves \cite{budzinski2022geometry,budzinski2023analytical, wrnn, magnasco2024inputdrivencircuitreconfigurationcritical} and complex spatiotemporal patterns \cite{budzinski2023theory, nwm} in nonlinear oscillator networks, demonstrating spatiotemporal `wave-like' dynamics are indeed beneficial for tasks requiring visual flow transformations.

In the context of memory, Karuvally et al. (2024)\cite{karuvally2024hidden} have demonstrated that even in simple recurrent neural networks with dense hidden-to-hidden connectivity matrices trained to solve memory retention tasks, it is possible to identify an alternative basis in which waves can be seen to propagate. However, once the connectivity of the recurrent connections is constrained to be local (e.g. through a convolution operation over the hidden state),  Keller et al. (2024)\cite{wrnn} demonstrated that the hidden layers converge on wave dynamics as the preferred way to propagate hidden activity in order to hold information in working memory for a long time. This work also demonstrates the clear benefit of the convolution-induced spatiotemporal inductive bias -- when compared with networks which did not possess such an inductive bias, the wave-based models trained orders of magnitude more quickly and solved memory tasks with significantly longer time horizons\cite{wrnn}. 
In the context of partially observed world state flows, it has been demonstrated that flow equivariant agents indeed are able to more quickly learn to represent partially observed dynamic environments, including having better memory for items that they see only briefly, and therefore are able to better predict future dynamics \cite{lillemark2026flowequivariantworldmodels}. %

Most surprisingly, however, our proposed perspective can be found to describe many of the design choices of state-of-the-art models that were designed without a prior notion of spatiotemporal dynamics or equivariance. For example, many modern computer-vision video models incorporate explicit `feature alignment' or `representation warping' modules that align past-frame features to the current frame (via optical-flow warping or learned deformable alignment) prior to temporal aggregation \cite{Zhu_2017_CVPR, Zhu_2017_ICCV, Tian_2020_CVPR, Zou_2021_CVPR, DBLP:conf/nips/ShiGXWYD22}. This heuristic has significantly improved accuracy in tasks such as video object detection and video restoration, and can be seen as a rough approximation to flow equivariance when the full group structure of the observed flow is unknown. Similarly, in recent years, a number of recurrent neural network architectures have emerged which have been shown to be competitive with the classical Transformer architectures on long-range sequence modeling for natural language generation. These include the Mamba \cite{gu2023mamba}, Hungry Hungry Hippos (H3) \cite{fu2023hungry}, xLSTM \cite{beck2024xlstm} and Griffin \cite{de2024griffin} architectures. Interestingly, all these state-space architectures include a convolution operation over sequence length at each layer, apparently inherited from the initial H3 framework of Fu et al.\cite{fu2023hungry} which proposed a loose intuition for why this may serve to remember tokens from the past.
The continued use of this architectural motif has been perpetuated as part of the empirical black-box of tricks required to reach strong performance with large recurrent neural networks; however, from the perspective outlined in this paper, we can now see that this motif is an instantiation of structured spatiotemporal neural dynamics. Specifically, this convolution operation is derived from a recurrent neural network defined as a `shift-SSM'\cite{fu2023hungry} which maintains a `shift' matrix as its recurrent connectivity matrix. This shift-SSM is identical to a special case of the wave-RNN of Keller et al.\cite{wrnn}, and can be derived as the finite difference approximation to the one-way one-dimensional wave equation -- analogous to the shift demonstrated for memory in Box \ref{fig:memory}. 
This congruence of findings presents evidence that a spatiotemporal perspective is not only relevant for understanding hypothetical solutions to memory storage but is also applicable to the near optimal solutions implemented in modern state-of-the-art artificial models.

Finally, we argue that even the autoregressive loop of generative Transformer architectures\cite{NIPS2017_3f5ee243} can be seen as instantiating a traveling wave which propagates backwards in time. As elaborated by Muller et al. (2024)\cite{muller2024transformers}, there is a clear connection between the extended `context' of a Transformer and the wave-field based memory seen in the aforementioned wave-RNNs\cite{wrnn}. At a high level, Transformers operate on entire sequences at once (often called a context), allowing them to learn long-range correlations by directly comparing elements of a sequence at different timesteps through a mechanism referred to as attention. In recurrent neural networks (such as the brain), however, there is no such explicit memory storage system, and therefore the fixed-size hidden state must be leveraged in order to store incoming inputs to allow for later cross-time comparisons. Muller et al. argue that traveling waves may serve exactly this `context' purpose. Karuvally et al. (2024)\cite{karuvally2024hidden} have further formalized this connection mathematically by showing that autoregressive Transformers may equivalently be viewed as wave-based models with the Transformer network itself operating as a non-linear boundary condition operating on the wave-field.

Overall, we believe these results point toward a functional convergence of artificial and biological neural networks. We present these models as a demonstration of the potential of our proposed perspective as a unifying conceptual framework, formalized by the language of equivariant neural network theory, which may help to interpret these findings and present a road map for promising directions of future research.

%% file: sections/7_Conclusion.tex
\section*{Discussion}
Early conceptual frameworks for computation in brains focused primarily on the properties of single neurons and their response to external stimuli. In light of modern developments, it is clear that brains are dynamical systems and cortical networks in particular are highly recurrent, with significant interactions between neighboring neurons\cite{rajan_dynamics}. As outlined in this article, the existence of cortical traveling waves leads to a new dynamic paradigm in neural computation that accounts for both the spatial organization of classical selectivity and prominently structured spatiotemporal dynamics. 

Our proposed spatiotemporal perspective on neural dynamics is a first step that uniquely brings together previously disparate ideas. From the viewpoints of mathematics, physics, and modern machine learning, spatiotemporal dynamics have a direct connection with the symmetries that inherently define our world. From the perspective of neuroscience, the structured spatial organization of selectivity may effectively represent these flow symmetries in a computationally useful way that is complementary to existing conceptual frameworks. 

As we have alluded to and argued for throughout this article, several predictions for neuroscience follow from this framework:
\begin{itemize}
    \item It should be possible to correlate stimulus flows with latent flows. The primary complication with testing this hypothesis is the overwhelming number of potential confounding internal `flows' that may be propagating simultaneously, reducing measurable signal. However, with proper controls and improved high resolution recording technology, such confounding factors are likely possible to eliminate. 
    \item Beyond stimulus flows, spatiotemporal dynamics should be measured to be in some way `yoked' to self-motion. It is known that motor signals are present throughout the cortex -- our framework suggests that we should be able to decode motion from the way neural activity \emph{transforms over time}, rather than simply its instantaneous presence. In other words, motion signals should serve to dynamically transform neural representations in a general manner rather than simply superimposing on top of them.
    \item It should be possible to observe `memory' as spatial waveforms of neural activity regardless of its cortical location (it should be translation invariant over the surface of the cortex). A decoder trained at one cortical location could be used at another location in the brain, instantiating a structured transformation operator between them. Evidence for this phenomenon was reported as early as 1934 by Adrian and Matthews\cite{adrian1934}, and by many others since then; however, without functional motivation, these reports have been ignored. We now have the computational tools and methods for analyzing spatiotemporal recordings to validate these ideas. 
    \item Finally, we should see a greater relationship between cortical organization of selectivity and spatiotemporal dynamics. There is emerging computational evidence that supports this \cite{Davis2024}, and more controlled developmental studies may solidify the role of spatiotemporal dynamics on structured selectivity in this regard.
\end{itemize}

The spatiotemporal perspective is a way to improve our interpretation of neural data and further enrich our understanding of the computational mechanisms underlying biological systems. Future work will refine and expand on these ideas, improving both our understanding of how biological neural systems encode and process information, and our ability to construct artificial neural systems with more natural representations\cite{ballard1999introduction}.

Ultimately, our core thesis can be summarized in the following four points: (I) In order to accurately represent input stimulus flows (`motion'), a recurrent neural network must have a corresponding flow of activity propagated through recurrent connections. (II) When there are no flow transformations acting on the input stimulus directly, recurrent flows of activity may still be used to model abstract `world state' flows, and thereby achieve structured working memory.  %
(III) Due to biological pressures, these latent flows are likely to take the form of roughly traveling-wave-like spatially organized dynamics. 
(IV) Due to the natural inclination of biological neural networks towards the production of such spatiotemporal dynamics, it is likely that they serve as an innate inductive bias for representing flows of the world.

\section*{Conclusions}

Spatiotemporal neural representations are not as intuitive a conceptual framework as the traditional one based on the receptive fields of single neurons.  Quantum mechanics was also not as intuitive as classical mechanics at the turn of the 20th century and introduced the duality between particles and waves.  We have assembled converging evidence from neuroscience and artificial neural networks that puts cortical waves center stage for this new perspective, analogous to how quantum mechanics introduced waves into our fundamental understanding of physics. The benefits of traveling waves range from flexibly handling movement symmetries and abstractions from the world, to holding information in working memory for extended periods of time, both properties which are currently handled by expensive mechanisms such as self attention in transformers. We suggest that with careful development and continued inquiry, the conceptual framework outlined here could elucidate some of the deep mysteries in both neuroscience and artificial intelligence.

\vspace{10 mm}